\documentclass[aps,prb,showpacs,onecolumn,preprintnumbers,amsmath,amssymb,amsfonts,floatfix]{revtex4-1}

\usepackage{graphicx, color,rotating, textcomp} 
\usepackage{dcolumn} 
\usepackage{bm} 
\newcommand{\sgn}[1]{\text{sgn}\left(#1\right)}
\newcommand{\rv}{\vec{r}}
\newcommand{\rvp}{\vec{r}^{\;\prime}}
\newcommand{\kv}{\vec{k}}
\newcommand{\qv}{\vec{q}}

\newcommand{\Imt}{\textrm{Im}}
\newcommand{\m}[1]{\mathbf{#1}}
\newcommand{\Nt}{\tilde{\mathcal{N}}}
\newcommand{\Mt}{\tilde{\mathcal{M}}}

\begin{document}

\title{Spectral footprints of impurity scattering in graphene nanoribbons}

\author{Anders~Bergvall}
\author{Tomas~L\"ofwander}
\affiliation{Department of Microtechnology and Nanoscience - MC2,
Chalmers University of Technology, SE-412 96 G\"oteborg, Sweden}

\date{\today}

%
%

\begin{abstract}

  We report a detailed investigation of the interplay between size
  quantization and local scattering centers in graphene nanoribbons,
  as seen in the local density of states. The spectral signatures,
  obtained after Fourier transformation of the local density of
  states, include characteristic peaks  that can be related to the transverse modes
  of the nanoribbon. In armchair ribbons, the Fourier transformed density of
  states of one of the two inequivalent sublattices takes a form
  similar to that of a quantum channel in a two-dimensional electron
  gas, modified according to the differences in bandstructure. After
  addition of the second sublattice contribution, a characteristic
  modulation of the pattern due to superposition is obtained, similar
  to what has been obtained in spectra due to single impurity
  scattering in large-area graphene. We present analytic results for
  the electron propagator in armchair nanoribbons in the Dirac
  approximation, including a single scattering center within a
  T-matrix formulation. For comparison, we have extended the
  investigation with numerics obtained with an atomistic recursive
  Green's function approach. The spectral signatures of the atomistic
  approach include the effects of trigonal warping.
  The impurity induced oscillations in the local density of states are
  not decaying at large distance in few-mode nanoribbons. 

\end{abstract}


\maketitle

%
%

\section{Introduction}

In graphene, scattering centers such as impurities, defects, adatoms,
and substrate inhomogeneities greatly influence the local electronic properties.\cite{CastroNeto:2009cl}
In some samples, the material quality is so high that a single or a few such scattering centers can
influence the whole device. This may degrade device function, but can also be taken advantage of
by making various sensing devices.\cite{Schedin:2007hs,Brar:2011iq}
Great attention has therefore been focused on understanding the influence
of scattering on the electronic properties of graphene.\cite{Peres:2010tn}

In this context, the scanning tunneling microscope (STM) is becoming
of increasing importance.\cite{Connolly:2010bc,Deshpande:2012jg}
By utilizing its various modes of operation, the STM can be used to map out
topography,  local density of states, local charge density, and more.
In this way, a variety of properties of graphene have been revealed.
A few examples include perturbations in the local density of states
around impurities\cite{Rutter:2007ep,Mallet:2007fg} or near step-edges in the substrate,\cite{Xue:2012hd}
charge puddle formation caused by molecules
trapped between graphene flakes and the  SiO$_2$ substrate,\cite{Zhang:2009ce}
and resistance caused by steps\cite{Ji:2011bla,Wang:2012tj} in the substrate or
multilayer regions\cite{Giannazzo:2012ib} in epitaxial graphene on silicon-carbide.

At the same time, encouraging progress has been achieved with fabrication
of graphene nanostructures. Top-down approaches include
nano-lithography,\cite{Han:2007bl} scanning probe methods,\cite{Biro:2010ez}
etching with metal nanoparticles along certain crystal directions,\cite{Campos:2009cd}
and utilization of the transmission electron microscope (TEM) to simultaneously image
and sculpture graphene.\cite{Girit:2009tz}
A bottom up approach based on chemical synthesis has also been demonstrated.\cite{Cai:2010bd}
Another approach involves unzipping of carbon nanotubes.\cite{Li:2008ht}
With that method, the theoretically predicted zero-energy (midgap)
edge states of nanoribbons with zigzag edges\cite{Fujita:1996vs,Nakada:1996us}
were directly mapped out by scanning tunneling spectroscopy (STS).\cite{Tao:2011bk}
Theory also predicts that by controlling the width and edges of nanoribbons, a bandgap can
be opened up at the Dirac point through quantum confinement
(see the review \onlinecite{Palacios:2010ep}).
With further progress it may soon become possible to study in much greater detail the interplay
between quantum confinement and impurity scattering in graphene nanoribbons.

Many theoretical studies of graphene nanoribbons have been reported in the literature,
see the collection of review articles in Ref.~\onlinecite{Raza}.
The effect of impurity scattering and the effects of edge
disorder on electron transport have been reported in several numerical
works. In an effort to simulate the typical experimental situation,
random disorder is included and the scaling behavior of resistivity
with length of the ribbon is studied, revealing different transport
regimes depending on ribbon width and disorder properties. Here, we go
back to the well defined problem of a single impurity in order to
study in detail the effects on the FT-LDOS.

In this paper we present results for the spectral signatures of a
local scattering center in graphene, taking into account quantum
confinement in a nanoribbon geometry.
This study generalizes the consideration of FT-LDOS of a single impurity in
bulk graphene\cite{PeregBarnea:2008ig,Bena:2008iw} to the case of nanoribbons.
We focus the analytic analysis
on armchair ribbons in the Dirac approximation (linearization around
the K-points in the graphene bandstructure), for which the
wavefunctions and propagators for clean ribbons are known, and solve
the impurity problem in a T-matrix formulation. Thereby we obtain the
electron propagator for an armchair nanoribbon including the effects
of a local scattering center. The Fourier transformed density of
states (FT-LDOS) is then obtained and explained in terms of scattering
processes of Dirac quasiparticles confined in the ribbon. We extend
the analysis to an atomistic tight-binding model of graphene,
utilizing a numerical recursive Green's function approach. The main
effect of going beyond the Dirac approximation is trigonal warping,
which shows up as a triangular distorsion of the FT-LDOS patterns.

For comparison we include an analysis of the FT-LDOS in a quantum ribbon
in a two-dimensional electron gas (2DEG). Many features of the FT-LDOS
patterns in graphene ribbons can be understood from the
somewhat simpler case of a 2DEG, and the new features special for
graphene can be highlighted. These include a more complicated
bandstructure due to the two inequivalent K-points, trigonal
warping, as well as interference effects due to the bipartite lattice
of graphene.

The outline of the paper is the following. In Section~\ref{Ch_FTSTS} we
discuss the Fourier transform scanning tunneling spectroscopy method and illustrate
the basic scattering processes at play in a nanoribbon.
In Section~\ref{Ch_2DEG} we present results for the FT-LDOS in a 2DEG quantum
channel. In Section~\ref{Ch_armchair} we report our results for the FT-LDOS in
an armchair graphene nanoribbon within the Dirac approximation and
compare with the 2DEG case. In Section~\ref{Ch_numerics} we present results of
numerical simulations of a tight-binding model, including also zigzag
nanoribbons as well as effects of edge disorder on the FT-LDOS. In Section
\ref{Ch_summary} we summarize the paper and give some conclusions and
an outlook. Most technical results of the analytic analysis have been
collected in the Appendices.

%
%

\section{Fourier transform  scanning tunneling spectroscopy}
\label{Ch_FTSTS}

A scattering center induces a perturbation of the local density of
states in its vicinity. For elastic scattering, the impurity scatters electrons between states
$\kv_1\rightarrow\kv_2$ with $\epsilon_{\kv_1}=\epsilon_{\kv_2}$, i.e. on
a contour of constant energy $E$. This leads to interference and a
wave pattern in the local density of states near the impurity with
wavevectors $\qv=\kv_2-\kv_1$. After Fourier transformation of the local
density of states $\rho(\rv,E)\rightarrow {\cal N}(\qv,E)$, the wave
vectors of the interference pattern are highlighted. The resulting
pattern in ${\cal N}(\qv,E)$ can then be used to infer the band dispersion
$\epsilon_{\kv}$. For instance, this has been done for metal surfaces.\cite{Anonymous:IEkMmlpc}
This method has also become a valuable tool for probing the properties
of high-$T_c$ superconductors.\cite{Balatsky:2006ce}

It is worth mentioning that, neglecting electron-electron interactions,
the interference patterns in the local
density of states discussed above are related to the Friedel
oscillations in the electron density $n(\rv)$ through integration over
energy including the Fermi-Dirac distribution function, $n(\rv)=-e\int
\rho(\rv,E)\,f(E)dE$.

By using the STM, the local density of states can be extracted as
function of energy by applying a finite voltage between tip and
sample, i.e. by employing scanning tunneling spectroscopy (STS). By
combining Fourier transformation with STS, the band dispersion can be
studied in the vicinity of the Fermi energy. This method has therefor
become a valuable spectroscopic tool sometimes called Fourier
transform scanning tunneling spectroscopy (FT-STS). In graphene, the
Fermi energy itself is tunable by a back gate voltage on the substrate
graphene is resting on. Thereby, FT-STS is potentially a valuable tool
for studies of graphene. Indeed,  experiment reproduce the
graphene bandstructure.\cite{Mallet:2007fg}

STS bears similarities with angle-resolved photo-emission (ARPES). STS
is ideal for spatially inhomogeneous systems, while ARPES relies on
large-area spatially homogeneous samples. Indeed, STS measures the
spatially resolved spectral function, i.e. local density of states $\rho(\rv;E)$,
while ARPES measures the momentum-space spectral function $A(\kv;E)$. By
generalizing STS to FT-STS, i.e. Fourier transforming
$\rho(\rv;E)\rightarrow {\cal N}(\qv;E)$, a spectroscopy has been introduced
that can be used to study materials, although we should remember
that ${\cal N}(\qv;E)$ is not equal to $A(\qv;E)$.

One advantage of FT-STS is the possibility to study nanoscale systems
with high spatial resolution. In this paper we will investigate the
consequences of quantum confinement on impurity scattering in 
graphene, as seen in FT-STS.

\begin{figure}
\includegraphics[width=\columnwidth]{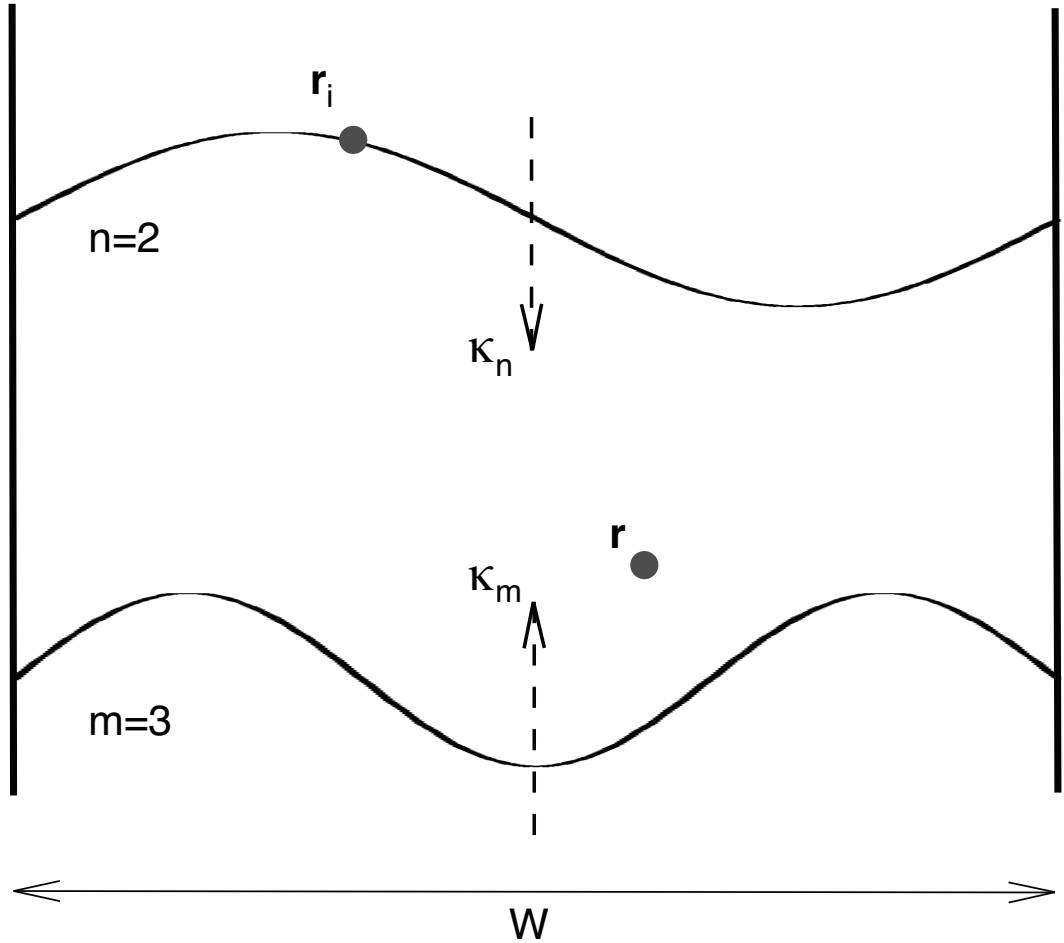}
\caption{An electron quasiparticle initially in mode $m$, with longitudinal
wave vector $\kappa_m$ may be backscattered by an impurity at $\rv_i$
into mode $n$. The local density of states at $\rv$ is changed due to interference
of the initial and final waves. This leads to an interference pattern around $\rv_i$.}
\label{Fig_handwave}
\end{figure}

In Fig.~\ref{Fig_handwave} we display a cartoon of a typical scattering
process that contributes to the correction to the local density of
states in a quantum ribbon with one impurity. For simplicity we here
discuss the situation in a 2DEG quantum channel. Quasiparticles
occupying for instance mode $m$, propagating in the positive
$y$-direction with wavenumber $\kappa_m$, passes the probing position
$\rv=(x,y)$, after which they can be backscattered by the impurity at $\rv_i$
into mode $n$ with wavenumber $\kappa_n$ and propagate back to the
probing position $\rv$. In this example we neglect evanescent modes
for simplicity. The contribution to the full propagator from this scattering
event will be proportional
to the free propagators before and after scattering and
the potential strength $\gamma$,
\begin{eqnarray}
\tilde G_{nm}(\rv,\rv;E) &=&
-i\frac{\mu}{\hbar^2}
\frac{e^{i\kappa_n|y_i-y|}}{\kappa_n}
\chi_n(x)\chi_n(x_i)\nonumber\\
&&\times\gamma
\left(-i\frac{\mu}{\hbar^2}\right)
\frac{e^{i\kappa_m|y-y_i|}}{\kappa_m}
\chi_m(x)\chi_m(x_i),\nonumber
\end{eqnarray}
where $\chi_n(x)=\sqrt{2/W}\sin(n\pi x/W)$ is the transverse
wavefunction in mode $n\ge 1$, and $\mu$ is the electron effective mass.
Taking into account multiple scattering by the impurity,
the potential strength $\gamma$ is replaced by a $T$-matrix.
When we take the imaginary part of the
propagator to get the local density of states, we get spatially
oscillating terms
\begin{eqnarray}
\propto \cos[(\kappa_n+\kappa_m)|y-y_i|],\nonumber
\end{eqnarray}
and
\begin{eqnarray}
\propto \sin[(\kappa_n+\kappa_m)|y-y_i|],\nonumber
\end{eqnarray}
since the $T$-matrix is a complex
number due to multiple scattering. After Fourier transformation, we
find peaks at $q_y=\pm(\kappa_n+\kappa_m)$ and at $q_x$ equal to
combinations of $n$ and $m$ times $\pi/W$. Thus, in a FT-STS picture
of a quantum channel, there will be a discrete number of peaks that
reflect the available modes. We also note that the Friedel
oscillations (neglecting electron-electron interactions) will at low temperature
oscillate without decay far from the impurity site.

To probe a 2DEG quantum channel with an STM in the way described here
will be challenging since the channel is typically hidden deep down in
a semiconducting heterostructure. Graphene, on the other hand,
is 100~\% surface and directly accessible.

%
%
\section{FT-LDOS: ribbon in a 2DEG}
\label{Ch_2DEG}

In this Section we improve the above discussion to the general case of
multiple scattering in a multimode 2DEG quantum channel of width $W$
with a single impurity scattering center at $\rv_i$. The results of this Section will be
referenced in the following Sections on graphene in order to highlight the distinguishing
features of confined Dirac quasiparticles.

Consider the probability amplitude for an electron in the channel to propagate from
one point $\rvp$ to another point $\rv$. For free propagation in mode $n$,  the amplitude 
is given by the free propagator (unperturbed Green's function), $g_n(\rv,\rvp;E)$.
In the presence of the impurity an electron initially in mode $m$ may be scattered into mode 
$n$. The effect of such an extra process will modify the propagator by adding a second term
\begin{equation}
\tilde{G}_{nm}(\rv,\rvp;E) = g_n(\rv,\rv_i;E)T(\rv_i;E)g_m(\rv_i,\rvp;E),
\end{equation}
so that the new Green's function will be
\begin{equation}
G_{nm}(\rv,\rvp;E) = g_n(\rv,\rvp,E)\delta_{nm} + \tilde{G}_{nm}(\rv,\rvp;E).
\end{equation}
The factor $T(\rv_i;E)$, see Eq.~(\ref{eqn:T_2deg}), includes multiple scattering by the impurity.
The full probability amplitude for propagation from $\rvp$ to $\rv$ is given by summing
over all mode indices
\begin{equation}
G(\rv,\rvp;E) = \sum_{nm}G_{nm}(\rv,\rvp;E).
\end{equation}

We may now proceed with the local density of states (LDOS). The correction to the LDOS 
by impurity scattering can be written as
\begin{eqnarray}
\tilde{\rho}(\rv;E) &=& -\frac{1}{\pi} \sum_{nm}\text{Im}\,\tilde{G}_{nm}(\rv,\rv;E)\nonumber\\
&=& -\frac{1}{\pi} \sum_{nm}\mathcal{K}_{nm}(E)\tilde{\rho}^x_{nm}(x;E)\tilde{\rho}^y_{nm}(y;E),
\end{eqnarray}
where the expressions for the factors $\mathcal{K}_{nm}(E)$, $\tilde{\rho}^x_{nm}(x)$,
and $\tilde{\rho}^y_{nm}(y;E)$ are given in Appendix~\ref{Appendix_2DEG}.
The Fourier transformed local density of states (FT-LDOS) can now be computed as
\begin{equation}
\Nt (\qv;E) = -\frac{1}{\pi} \sum_{nm}\mathcal{K}_{nm}(E) \Nt^x_{nm}(q_x)\Nt^y_{nm}(q_y;E),
\label{Eq_FTLDOS}
\end{equation}
where
\begin{equation}
\label{eqn:Nqx}
\begin{split}
\Nt^x_{nm}(q_x) &= \sum_{l=-\infty}^\infty \delta\left(\frac{q_x}{\pi}-\frac{l}{W}\right)\int_{-W}^W\frac{dx}{2W}\;e^{-i\frac{\pi}{W}l x}\tilde{\rho}^x_{nm}(x) \\
& = \sum_{l=-\infty}^\infty \delta\left(\frac{q_x}{\pi}-\frac{l}{W}\right)\Nt^x_{nm}(l)
\end{split}
\end{equation}
and
\begin{equation}
\Nt^y_{nm}(q_y;E) = \int_{-\infty}^\infty \frac{dy}{2\pi} e^{-iq_y y}\tilde{\rho}^y_{nm}(y;E).
\end{equation}
The function $\tilde{\rho}^x_{nm}(x)$, originally defined on the interval $[0,W]$,
is extended to $[-W,W]$ and assumed to be even with respect to the origin.
Due to the finite width, $2W$, of the integration interval, the spectral $x$-component 
is fixed to be integer multiples of $\pi/W$. This is a trick to be able to resolve the minimum change of transverse momenta, 
$\pi/W$, when scattering between two different modes.

It is important to realize that both propagating and evanescent modes play a role in
this scattering problem. The longitudinal momentum is
$\kappa_n = \sqrt{2\mu E/\hbar^2 - (n\pi/W)^2}$, 
where $\mu$ is the electron mass and $n\ge 1$ is the integer mode index.
At the bottom of a subband, $\kappa_n\rightarrow 0$, and the evanescent mode
extends far from the impurity and play an important role. On the other hand, for energies
far from any subband bottom, the local density of states is only affected by the evanescent
mode in a small region near the impurity. In the discussion of the FT-LDOS we can 
then safely neglect evanescent modes in the sums in Eq.~(\ref{Eq_FTLDOS}).
The evanescent modes are still taken into account in the scattering processes
at the impurity through the T-matrix equation, where intermediate modes can be evanescent,
while initial and final modes are propagating. In all of our numerical calculations, we include 10 evanescent modes. Adding even
more evanescent modes does not qualitatively change our results.
As have been shown, a delta-shaped impurity with a finite number of 
evanescent modes will model an s-like scatterer.\cite{2000PhRvB..61.5632B}
We can now find the different components of the FT-LDOS to be
\begin{equation}
\label{eqn:Nl}
\Nt_{nm}^x(l) = \frac{1}{2W}\left(\delta_{l,n-m} + \delta_{-l,n-m} - \delta_{l,n+m} - \delta_{-l,n+m}\right)
\end{equation}
and
\begin{equation}
\Nt_{nm}^y(q_y) = \frac{e^{-iy_iq_y}}{2\pi}\left[S^y(\kappa_n + \kappa_m + q_y) + S^y(\kappa_n + \kappa_m - q_y)\right],
\end{equation}
where
\begin{equation}
\label{eq:Sy_2DEG}
S^y(a) = \lim_{\epsilon \rightarrow 0^+} \frac{\sigma_p \epsilon - a(1/\gamma + \sigma_e)}{\epsilon^2 + a^2}
\end{equation}
and where $\sigma_{p/e}$ are positive, $\vec{q}$-independent constants defined in Eq. ~(\ref{eqn:sigmas_2deg}). 

The factor $\mathcal{K}_{nm}(E)$ is given by
\begin{equation}
\mathcal{K}_{nm}(E) = \frac{1}{(1/\gamma + \sigma_e(E))^2 + \sigma_p^2(E)} \left(\frac{\mu}{\hbar^2}\right)^2 \frac{\chi_n(x_i) \chi_m(x_i)}{\kappa_n(E) \kappa_m(E)}
\end{equation}
and depends on the scatterer strength $\gamma$ and the transversal wave 
functions $\chi_n(x) = \sqrt{2/W}\sin(n\pi x/W)$.

\begin{figure}
\includegraphics[width=0.49\columnwidth]{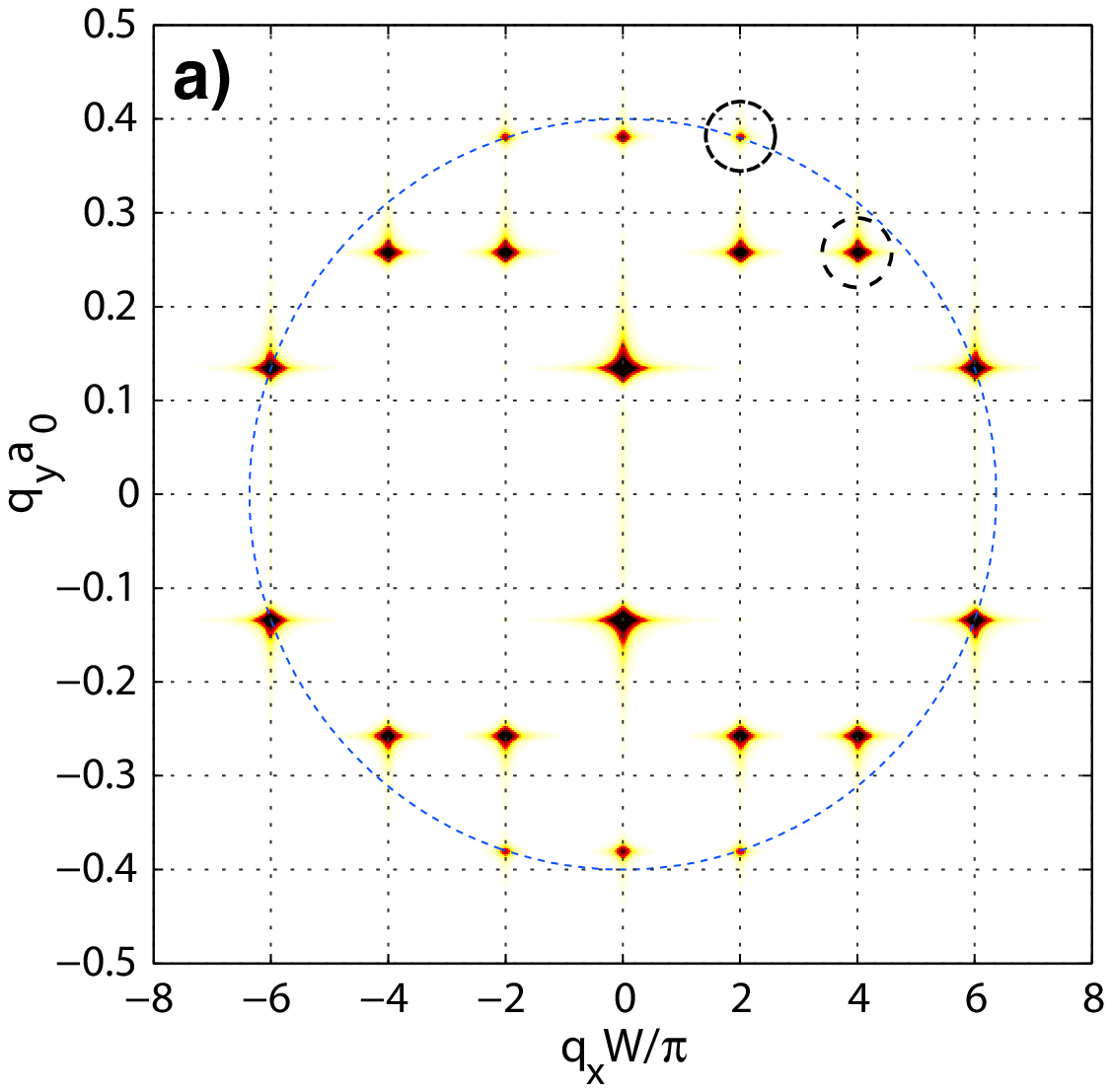}
\includegraphics[width=0.49\columnwidth]{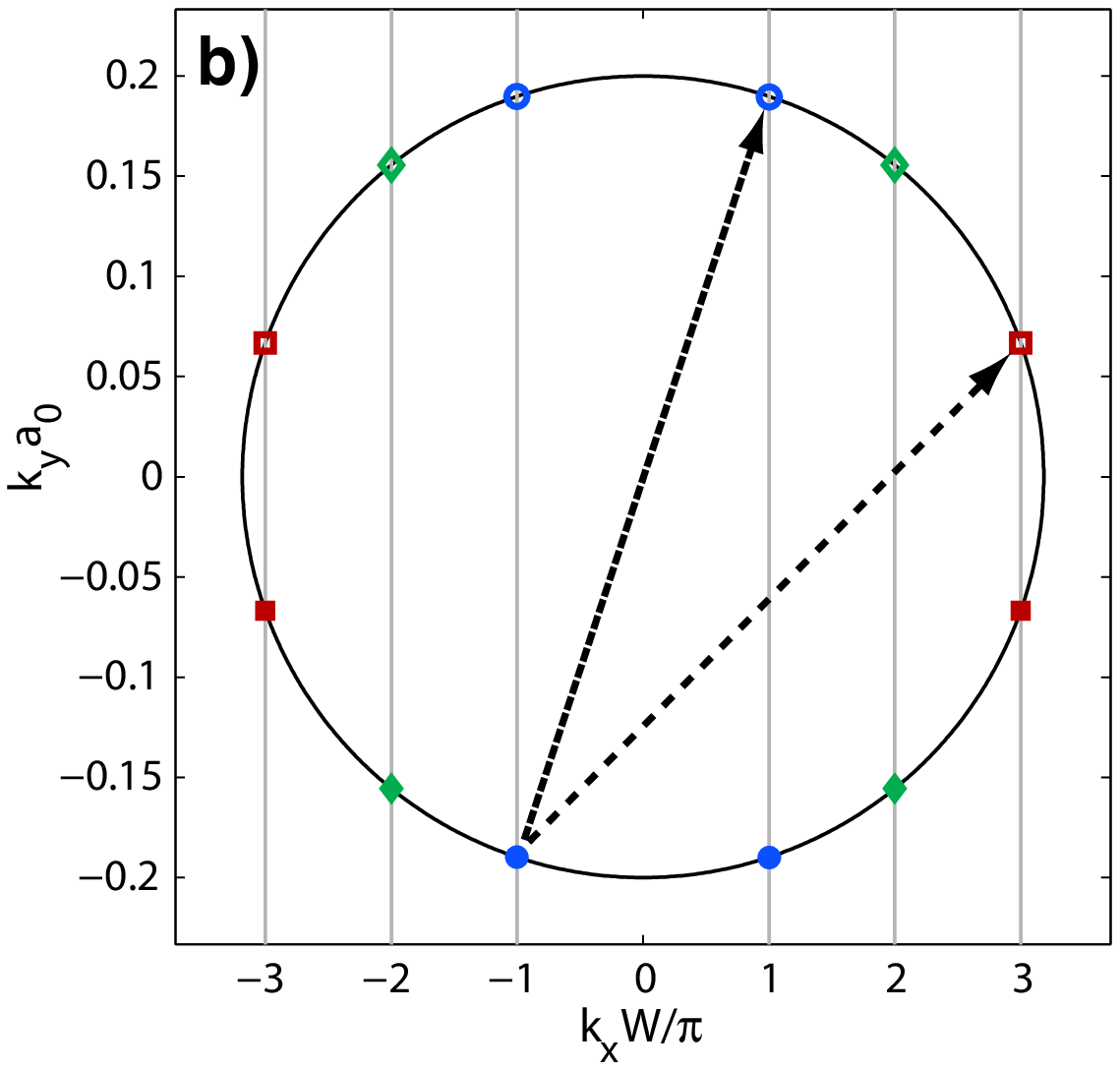}
\caption{(a) FT-LDOS of a 2DEG ribbon of width $W = 50a_0$ at energy $E = 0.025\tau$,
with three open (propagating) modes.
(b) Energy contour of the 2DEG dispersion. The symbols indicate the allowed
momentum values between which scattering can potentially take place
(circles, diamonds and squares corresponds to $n=1$, $n=2$ and $n=3$ respectively).
The two arrows in figure (b) illustrates two possible scattering
processes that gives rise to the two encircled dots in figure (a).
The impurity is placed at $x_i = W/2$ and the $n=2$ subband is not
scattered by the impurity because the impurity has been located at a node of
the corresponding transverse wavefunction.}
\label{fig:2DEG_ftldos_5}
\end{figure}

Together, these components give rise to a number of 
selection rules that govern the modification of the FT-LDOS 
by impurity scattering. To illustrate, we select 
a narrow channel ($W = 50a_0$, where $a_0$ defines the unit length) 
and a low energy ($E = 0.2\tau$, where $\tau$ defines the unit energy), 
such that only a total of three propagating modes are open. 
The scattering FT-LDOS $|\Nt(\vec{q};E)|$ for the case of the impurity in the middle
of the ribbon ($x_i = W/2, y_i = 0$) is displayed in Fig.~\ref{fig:2DEG_ftldos_5}(a).
Since the scattering is elastic, energy conservation requires that 
the transverse and longitudinal momenta, both before and after scattering,
satisfy the relation $2\mu E/\hbar^2 = k_x^2 + \kappa_n^2(E)$, 
which is the circle shown in Fig.~\ref{fig:2DEG_ftldos_5}(b).
In the channel, the transverse momentum is
quantized, $k_x \rightarrow k_n = n\pi/W$, and the only allowed momentum 
values between which the electrons can scatter are indicated by 
dots and squares on this circle. The FT-LDOS is therefor
non-zero only at a few, finite number of $\vec{q}$-points. All of
of these points lie inside the dotted circle of radius 
$2\sqrt{2\mu E/\hbar^2}$ shown in Fig.~\ref{fig:2DEG_ftldos_5}(a).

The factor $\mathcal{K}_{nm}(E)$ will be non-zero only if the transverse wavefunctions 
of mode $n$ and $m$ have a finite overlap at the position of the impurity. 
Since we have positioned the impurity at $x_i = W/2$, $\mathcal{K}_{nm}(E)$ will in this example
be non-zero only if $n$ and $m$ are both odd integers since all 
the wavefunctions with even indices will have a node at $x = x_i$. Thus, modes
with even number $n$ are not scattered by the impurity in this example.

To understand the exact locations of the $\vec{q}$-points, we start by looking 
at the case $q_x = 0$ (i.e., $l = 0$). Since all mode indices have to be odd, 
the term $\Nt^x(l=0)$ will be non-zero only when $n = m$, i.e., when ($n = 1, m = 1$) 
or when ($n = 3, m = 3$). This tells us that the points along $q_x = 0$ are 
all due to intraband scattering. The factor $\Nt^y(q_y;E)$ peaks when 
$q_y = \pm 2|\kappa_1|$ or when $q_y = \pm 2|\kappa_3|$. These are the four 
points we see along the line $l = 0$.

When $q_x = \pi/W$ ($l = 1$), at least one of the indices $n$ and $m$ will be even,
and the factor $\mathcal{K}_{nm}(E)$ is zero. This is why we see no bright points along this line.
This also happens for $l=3$ and $l=5$.

Along the line $q_x = 2\pi/W$, we have that $\Nt^x(l=2)$ is non-zero 
only when ($n = 1, m = 1$), ($n = 1, m = 3$) or when ($n = 3, m = 1$). The factor 
$\Nt^y(q_y;E)$ peaks at $q_y = \pm 2|\kappa_1|$ or when $q_y = \pm |\kappa_1 + \kappa_3|$, 
and we see that we have spots at these locations along $l = 2$ in the figure.

At $l = 4$ we must have $(n = 3, m = 1)$ or $(n=1, m=3)$,
which tells us that $q_y = \pm|\kappa_3+\kappa_1|$.
At $l = 6$, we must have ($n = 3, m = 3$) and $q_y = \pm2|\kappa_3|$.
A similar argument can be made for $l < 0$, and we can 
therefor say exactly which scattering processes contribute to each dark spot in 
Fig.~\ref{fig:2DEG_ftldos_5}(a). 

\begin{figure}
\includegraphics[width=\columnwidth]{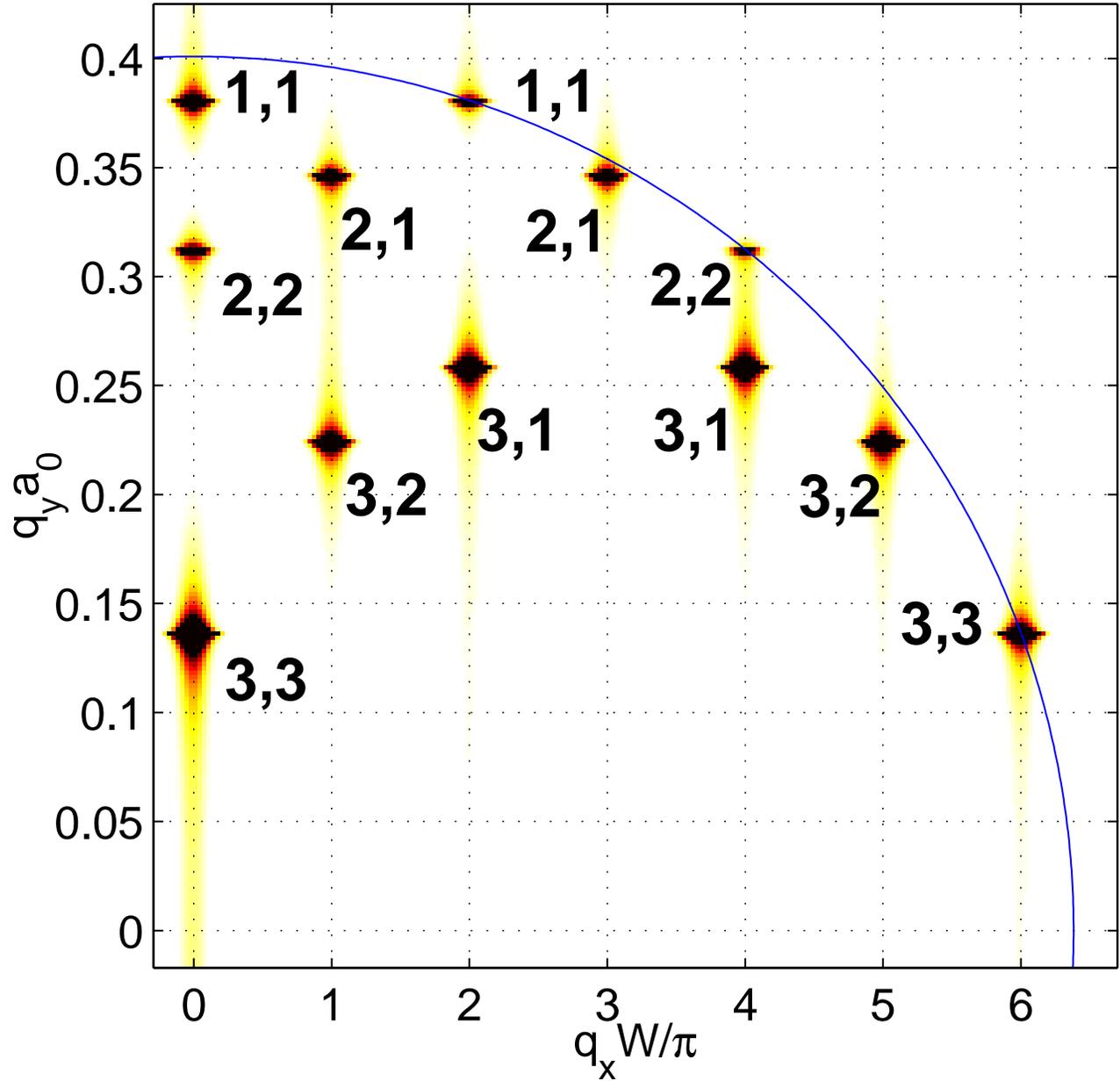}
\caption{The first quadrant of the FT-LDOS of the same ribbon as in Fig.~\ref{fig:2DEG_ftldos_5},
but with the impurity placed at $x_i = 2W/7$. The numbers indicate the modes that
gives rise to the different points.}
\label{fig:2DEG_ftldos_numbered}
\end{figure}

If the impurity is not located exactly at the middle of the ribbon, the even subbands will also be
part of the scattering process. This is illustrated in Fig.~\ref{fig:2DEG_ftldos_numbered}, where
we have numbered the subband transitions corresponding to each bright point.

\begin{figure}
\includegraphics[width=0.49\columnwidth]{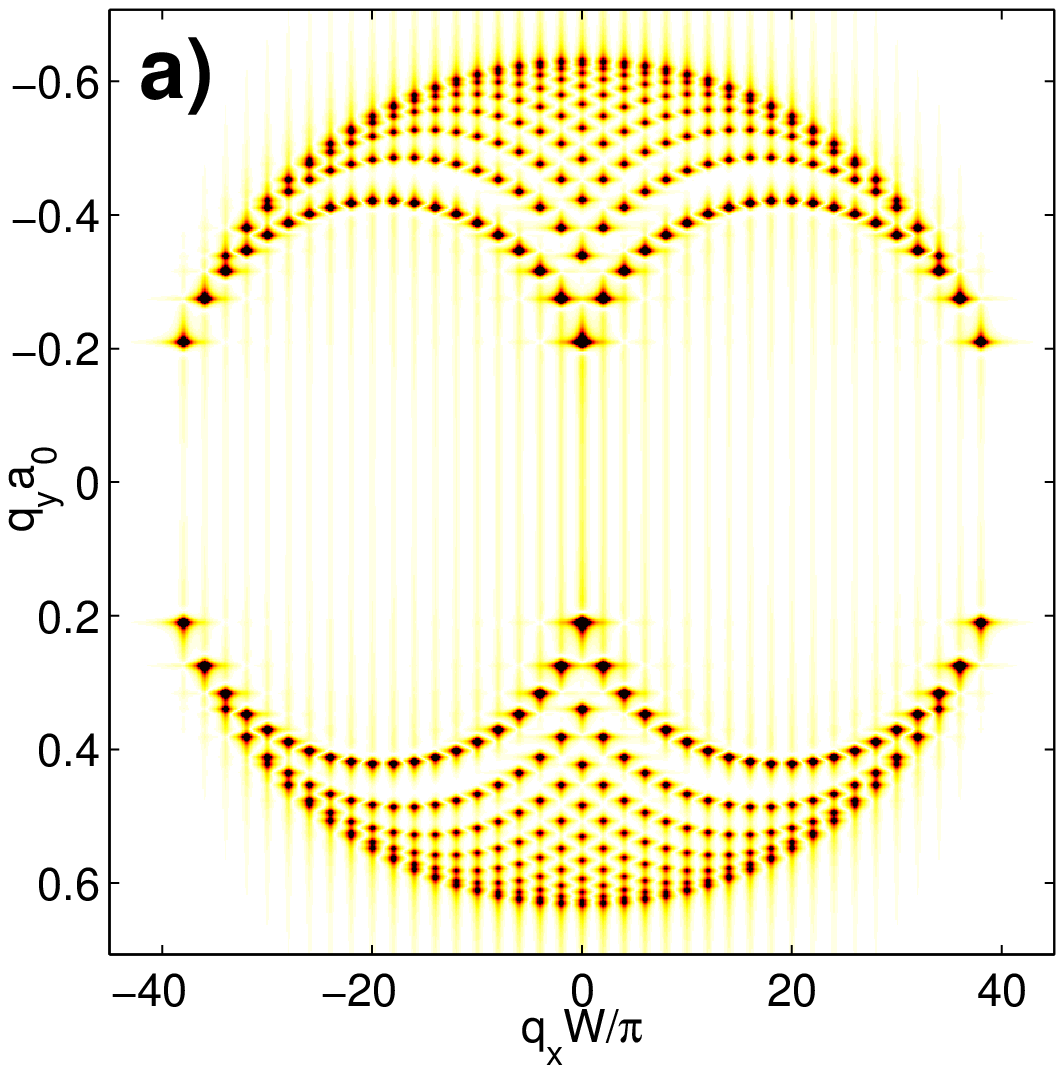}
\includegraphics[width=0.49\columnwidth]{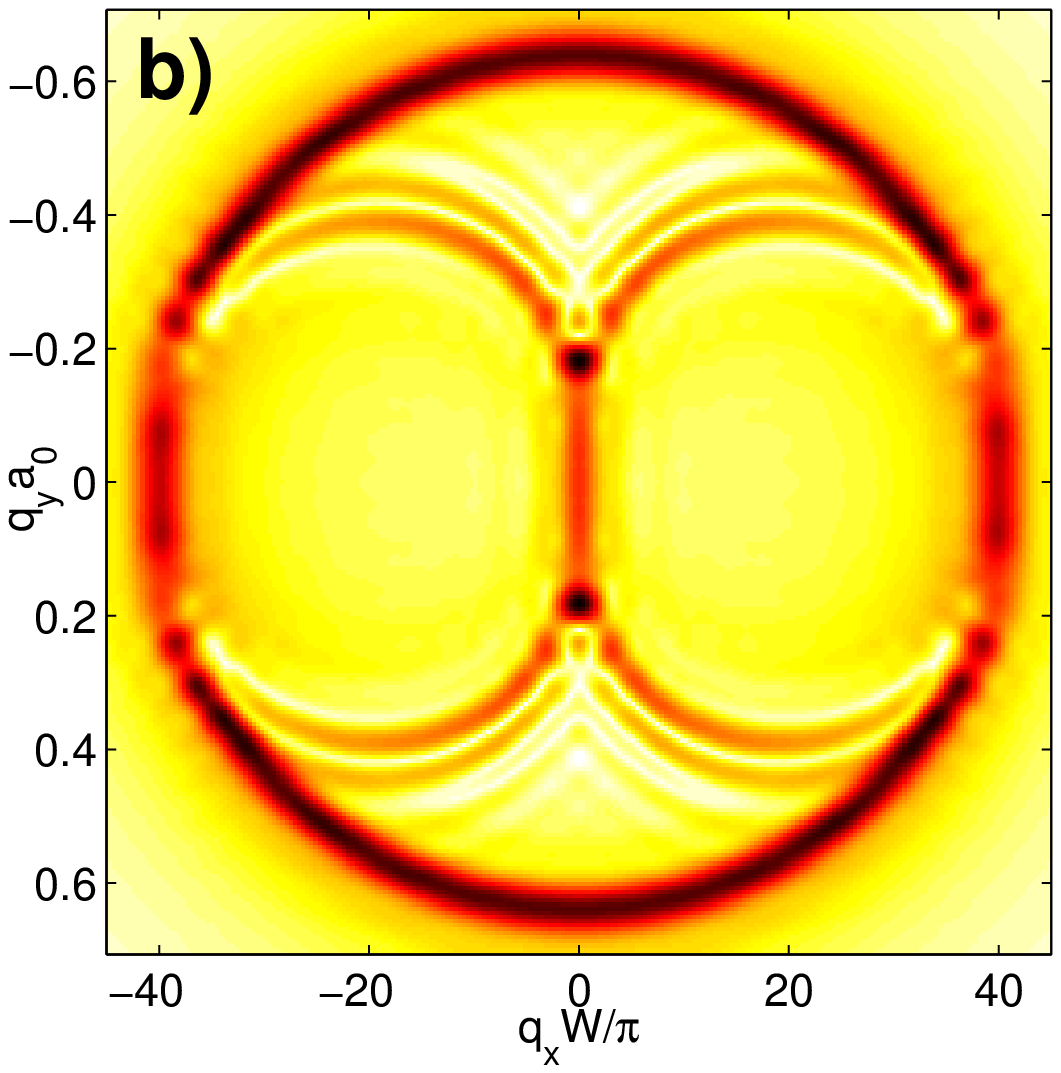}
\caption{(a) Analytical FT-LDOS of a 2DEG ribbon of width $W = 200a_0$ and energy 
$E = 0.05\tau$. (b) FT-LDOS taken from a numerical tight-binding simulation 
of a ribbon of width $W = 200a_0$ and energy $E = -3.95\tau$. The energies are adjusted 
such that each ribbon has a total of 20 propagating modes open.} 
\label{fig:ftldos_2degtb}
\end{figure}

In Fig.~\ref{fig:ftldos_2degtb}, we show the result for a wider ribbon calculated 
both analytically and by doing a recursive tight-binding simulation. 
The parameters are adjusted such that both cases have 20 propagating modes open,
and we see that the main features of our analytical calculation and the numerical simulation
coincide.

%
%

\section{FT-LDOS: ribbons of graphene}
\label{Ch_armchair}
The procedure of calculating the effect of a single impurity on the local density of states
in a graphene armchair  nanoribbon (AGNR) much follows that used for the 2DEG case.
Due to the bipartite structure of the graphene honeycomb lattice,
the propagator $\tilde{\mathbf{G}}(\rv,\rvp;E)$ is a 2x2 matrix in sublattice space, denoted by A and B. 
We therefor start by finding the impurity contribution to the local density of states
on each sublattice. The resulting
expressions for the A- and B-sublattice LDOS can be written as
\begin{equation}
\begin{split}
\tilde{\rho}^{A/B}(\vec{r};E) 
&= -\frac{1}{\pi}\sum_{dc}\sum_{nm} \text{Im}\;\tilde{G}^{AA/BB}_{dncm}(\vec{r},\vec{r};E) \\
&= -\frac{1}{\pi}\sum_{dc}\sum_{nm} \mathcal{K}^{A/B}_{dncm}(E)\tilde{\rho}^x_{nm}(x;E)\tilde{\rho}^{(A/B)y}_{dncm}(y;E), 
\end{split}
\end{equation}
where $\tilde{G}^{AA/BB}_{dncm}(\rv,\rv;E)$ are the two diagonal components of the 
propagator matrix $\tilde{\mathbf{G}}_{dncm}(\rv,\rv;E)$. The summation over 
the variables $c$ and $d$ are added to account for scattering between different sets of 
non-equivalent Dirac cone pairs $\vec{K}^{\pm}_c$ and $\vec{K}^{\pm}_d$. A further 
elaboration on this is found in Appendix~\ref{Appendix_armchair}, together with 
derivations of the expressions for $\tilde{\rho}^x_{nm}(x;E)$ and $\tilde{\rho}^{(A/B)y}_{dncm}(y;E)$.
As discussed in section~\ref{Ch_2DEG}, we only need to sum over propagating incoming and
final transverse modes, labeled by $m$ and $n$.

We compute the FT-LDOS on each sublattice as
\begin{equation}
\label{eq:M}
\tilde{\mathcal{M}}^{A/B}(\vec{q};E) 
= -\frac{1}{\pi}\sum_{nm}\tilde{\mathcal{M}}^x_{nm}(q_x;E) \tilde{\mathcal{M}}_{nm}^{(A/B)y}(q_y;E),
\end{equation}
%
The total FT-LDOS is found as a superposition of the two sublattices
\begin{equation}
\label{eqn:Ntot_agnr}
\tilde{\mathcal{M}}(\vec{q};E) = \tilde{\mathcal{M}}^A(\vec{q};E) + e^{-i a_0 q_y}\tilde{\mathcal{M}}^B(\vec{q};E), 
\end{equation}
where the extra phase-shift is introduced since the two sublattices are spatially separated by the carbon-carbon 
distance $a_0$ in the y-direction.

Since the transverse wavefunctions, $\chi_{n}(x) = \sqrt{1/W}\sin(n\pi/Wx)$, in our AGNR only differ from those 
of the 2DEG by a factor of $1/\sqrt{2}$, we have that $\tilde{\mathcal{M}}^x_{nm}(q_x;E) = \tilde{\mathcal{N}}^x_{nm}(q_x;E)/2$,
as defined in Eqns. (\ref{eqn:Nqx}) and (\ref{eqn:Nl}). 

The longitudinal FT-LDOS expressions for each sublattice are given by
\begin{equation}
\label{eqn:My}
\begin{split}
\tilde{\mathcal{M}}^{(A/B)y}_{nm}(q_y;E) 
&= \frac{e^{-iq_yy_i}}{2\pi}\sum_{c=1}^3 \mathcal{K}^{A/B}_{cncm}(E)\left[S^{(A/B)y}(\Delta_{cncm}(E) - q_y)^* + S^{(A/B)y}(\Delta_{cncm}(E) + q_y)\right] \\
&+ \frac{e^{-iq_yy_i}}{2\pi}\sum_{d=1}^2\sum_{c=d+1}^3 \mathcal{K}^{A/B}_{dncm}(E)\left[S^{(A/B)y}(\Delta_{cndm}(E)- q_y)^* + S^{(A/B)y}(\Delta_{dncm}(E)-q_y)^* \right.\\
&\left. \quad \quad \quad \quad \quad \quad \quad \quad \quad \quad \quad \; + S^{(A/B)y}(\Delta_{cndm}(E) + q_y) + S^{(A/B)y}(\Delta_{dncm}(E)+q_y) \right],
\end{split}
\end{equation}
where $\Delta_{dncm}(E) = \text{sgn}(E)[\kappa_{dn}(E) + \kappa_{dm}(E)] + K^y_d - K^y_c$.
The two $\vec{q}$-independent constants are found to be
\begin{equation}
\mathcal{K}^A_{dncm}(E) = \frac{1}{\left[1/\gamma + \sigma_e(E)\right]^2 + \sigma_p^2(E)}
\left(\frac{|E|}{v_f^2}\right)^2\frac{\chi_n(x_i)\chi_m(x_i)}{\kappa_{dn}(E)\kappa_{cm}(E)}
\end{equation}
and $\mathcal{K}^B_{dncm}(E) = -(v_f/|E|)^2 \mathcal{K}^A_{dncm}(E)$.
The transverse and longitudinal momenta are now cone set dependent, and changes to 
$k_{dn} = n\pi/W - K_{dx}$ and $\kappa_{dn}(E) = \sqrt{(E/v_f)^2 - k_{dn}^2}$ respectively.
Here, $S^{Ay}(a) = S^y(a)$ is the same function as used 
in the 2DEG case and defined in Eq.~(\ref{eq:Sy_2DEG}), and $S^{By}(a) = f_{dncm}(E) S^{y}(a)$ where
$f_{dncm}(E) = [-k_nk_m + \kappa_n(E)\kappa_m(E)] + i\text{sgn}(E)[k_n\kappa_m(E) + k_m\kappa_n(E)]$.
The $S$-terms for the AGNR A-lattice have the exact same form as the corresponding terms in the 2DEG,
while the B-lattice terms are scaled by a complex mode dependent prefactor. This is a consequence of our
choice of impurity potential: an impurity fully localized on one A-atom, see Eq.~(\ref{eq:Vgr}).

\begin{figure}
\includegraphics[width=0.5\columnwidth]{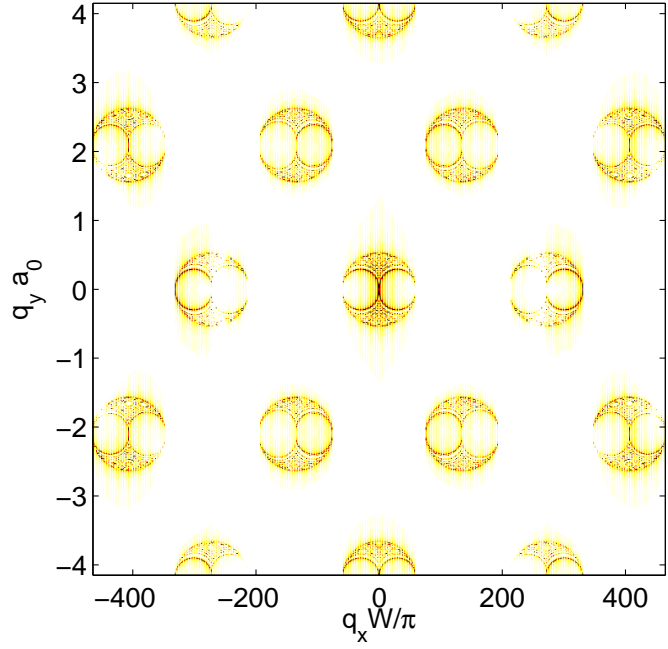}
\caption{Analytic FT-LDOS, $\Mt(\vec{q};E)$, of an AGNR having width $W \approx 50$, $E = 0.4|\tau|$, $\gamma = 10|\tau|$ and 60 
propagating channels.}
\label{fig:ftldos_full}
\end{figure}

In Fig.~\ref{fig:ftldos_full}, we plot $|\tilde{\mathcal{M}}(\vec{q};E)|$ for a semiconducting AGNR of width $W \approx 50$ nm, 
where $E=0.4|\tau|$ such that 60 channels are propagating. 
The positions, outer shapes, and sizes of the circular features (of radius $2E/v_f$)
are the same as those found when studying a sharp impurity in bulk graphene, and are due to the graphene bandstructure.
In addition to the bulk graphene features, we also see 
an added rich inner structure due to transverse confinement in the nanoribbon. Each peak corresponds to scattering processes
that change the transverse momentum by integer multiples of $\pi/W$, and change the
longitudinal momentum  such that the arguments of at least one of the many $S^y$-terms in Eq.~(\ref{eqn:My}) vanishes.

\begin{figure}
\includegraphics[width=0.5\columnwidth]{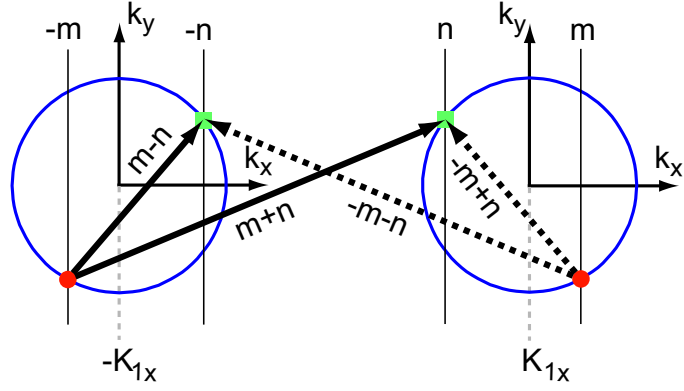}
\caption{Schematic picture of one possible scattering process in AGNR's.
Here, the electron (initially in mode $m$, represented by the red dots), is scattered into
mode $n$ (green squares). The FT-LDOS will be finite at the $\vec{q}$-values illustrated
by the solid ($q_x = (m\pm n)\pi/W$) and dotted ($q_x = -(m\pm n)\pi/W$) arrows.}
\label{fig:agnr_proc}
\end{figure}

A schematic illustration of one such scattering process is shown in
Fig.~\ref{fig:agnr_proc}. An electron, initially in cone pair $c=1$ and mode $m$,
is described by a plain wave with momentum $-\kappa_{1m}$
in the longitudinal direction, and a superposition of two plain waves with momenta $\pm k_{1m}$ in the transverse direction 
(see the lower red dots). After scattering (within the same cone pair) to mode $n$,
the momenta are changed to $\kappa_{1n}$ and $\pm k_{1n}$
in the longitudinal and transverse directions respectively (see the upper green squares).
A Fourier transform of the LDOS is proportional to a product of the electron wavefunction before and after the scattering event,
where each wavefunction is a linear combination of two transverse parts.
The FT-LDOS will therefor be finite at the $\vec{q}$-values corresponding to the four arrows shown in the figure.
Here, $q_y = |\kappa_{1n}+\kappa_{1m}|$, and $q_x = (m\pm n)\pi/W$ (solid arrows) or $q_x = -(m\pm n)\pi/W$ (dotted arrows).
When scattering to a different cone pair $d\neq c$, we instead have
$q_y = |\Delta_{dn1m}|$, see Eq.~(\ref{eqn:My}).

\begin{figure}
\includegraphics[width=0.49\columnwidth]{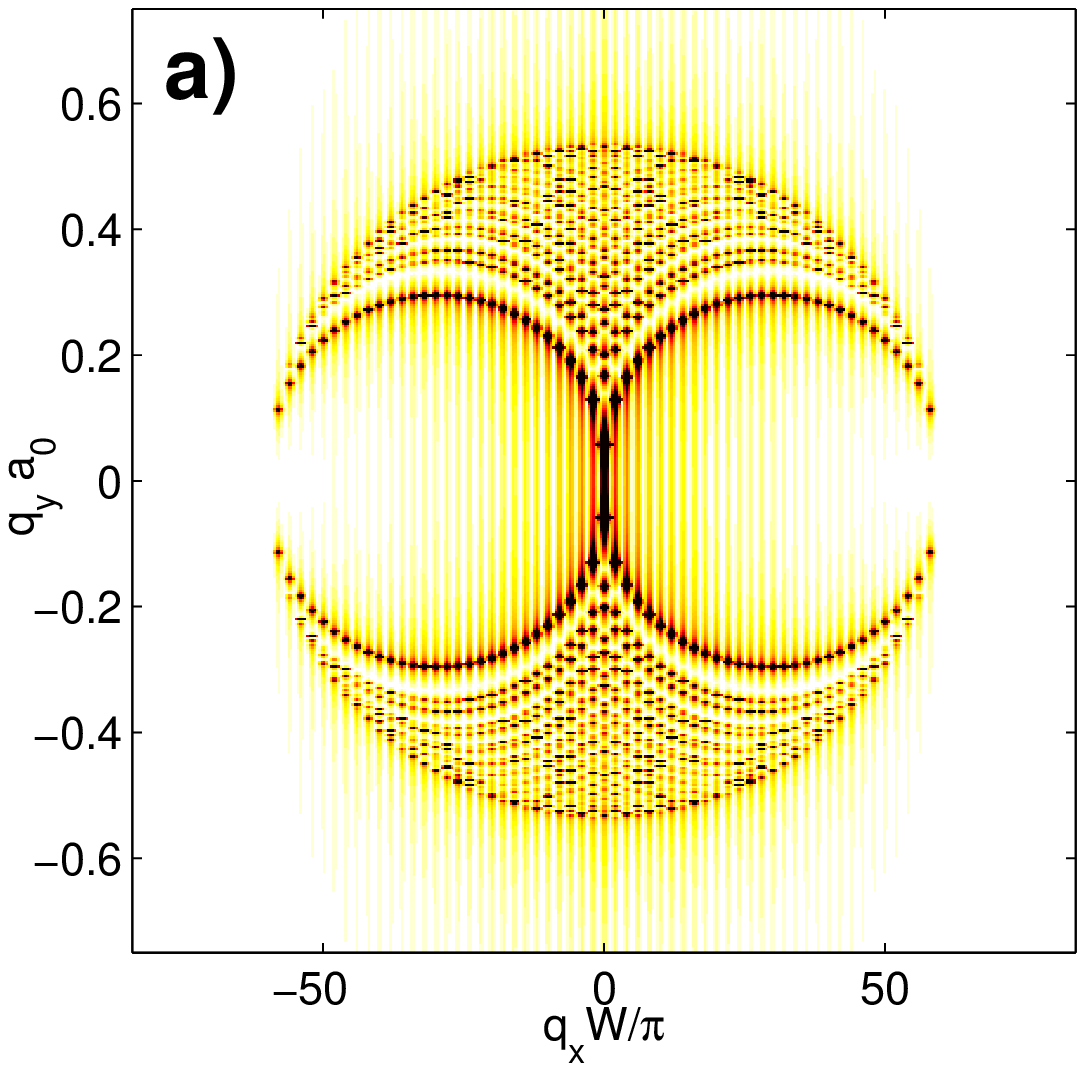}
\includegraphics[width=0.49\columnwidth]{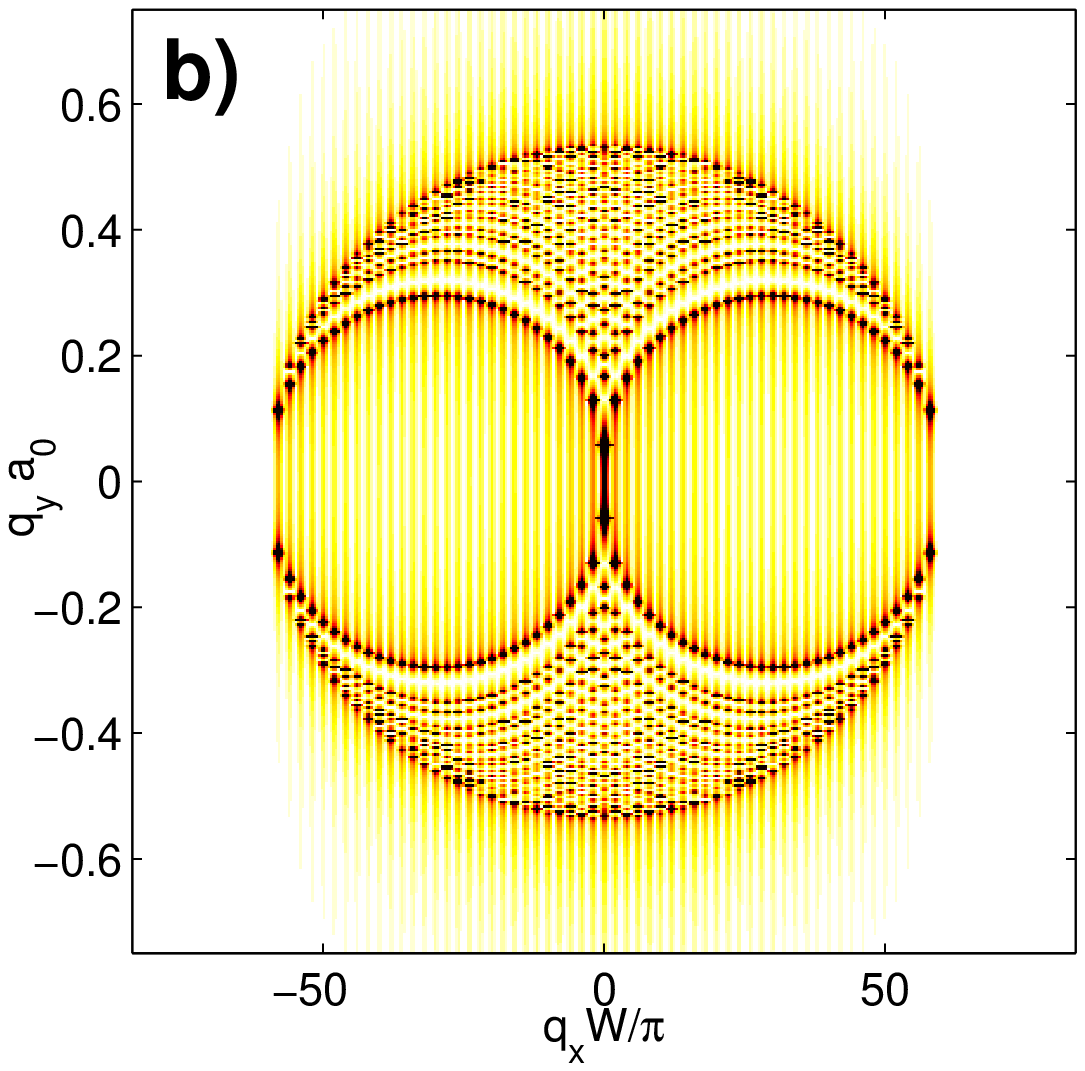}
\caption{Zoom-ins of the central circular feature of Fig.~\ref{fig:ftldos_full} showing (a) the combined A- and B-lattice 
contributions and (b) the A-lattice alone. When adding the two lattices together, the outer circle is attenuated by destructive 
interference between the two lattice contributions, $\Mt^{(A/B)y}(q_y;E)$, to the total FT-LDOS.}
\label{fig:ftldos_middle}
\end{figure}

When we zoom in on the circular 
feature in the middle [shown in Fig.~\ref{fig:ftldos_middle}(a)] we see that the outer ring of non-vanishing $\vec{q}$-points in 
$|\tilde{\mathcal{M}}(\vec{q};E)|$ appears to be attenuated compared with what is seen on e.g. the A-lattice alone 
[$|\tilde{\mathcal{M}}^A(\vec{q};E)|$ shown in Fig.~\ref{fig:ftldos_middle}(b)]. This is due to destructive interference when
adding the A- and B-lattice FT-LDOS contribution together, as done in Eq. (\ref{eqn:Ntot_agnr}). The $\vec{q}$-points on the outer 
circle comes from scattering processes which maximimize the change in momenta while still scattering within the same cone pair, 
i.e., where $d=c$ and $k_{cm} \rightarrow -k_{dn}$ and vice versa. In this case, we have that $f_{dncm} = (E/v_f)^2$ which tells 
us that $\Mt^{By}_{nm}(q_y;E) = -\Mt^{Ay}(q_y;E)$ so that when the phase factor $e^{-iq_ya}$ in Eq. (\ref{eqn:Ntot_agnr}) is close 
to unity, the contributions from the A- and B-lattice will cancel each other out. Similar cancellations may be seen in Fig 
\ref{fig:ftldos_full}, e.g. in the circular features to right and left of the central one. For other processes and $\vec{q}$-values, 
the interference between the two lattice contributions may not play an important role, or we might have constructive interference 
instead.

\begin{figure}
\includegraphics[width=0.49\columnwidth]{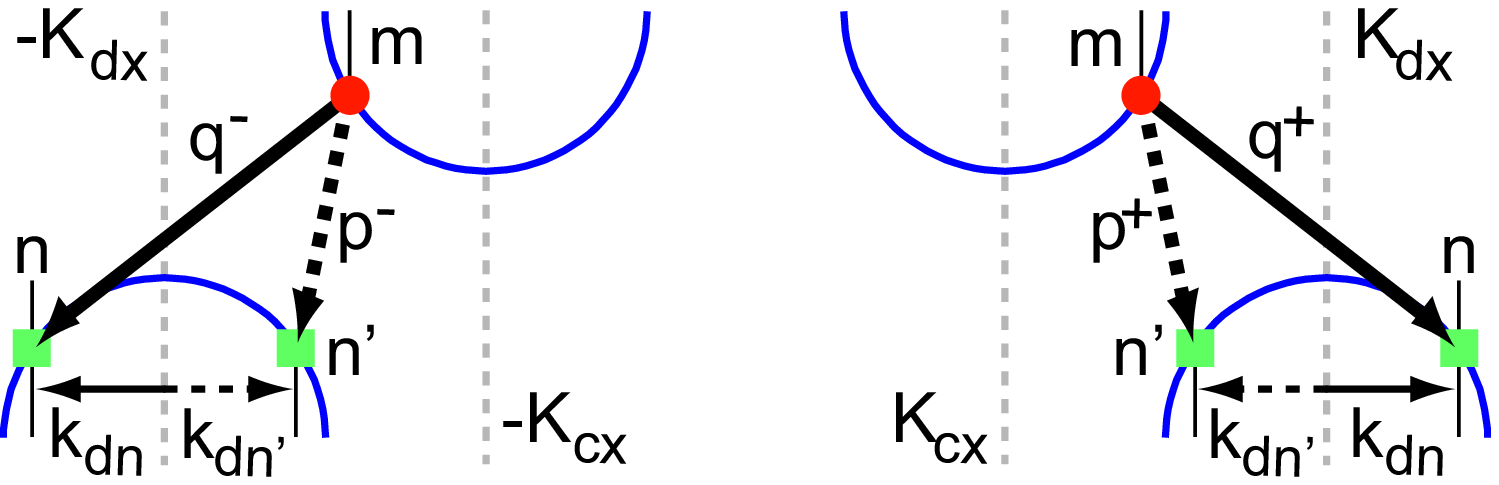}
\caption{Schematic picture of scattering processes conserving mirror symmetry with respect to $q_x = 0$ (solid arrows) and
with respect to $q_x = \pm K_{dx}$ (solid + dotted arrows).}
\label{fig:agnr_proc2}
\end{figure}

The FT-LDOS is left-right mirror symmetric around the line $q_x = 0$, see Fig.~\ref{fig:ftldos_full}.
This symmetry appears because for every process adding a
component $\vec{q}^+$ in the FT-LDOS, there is another process adding a component
$\vec{q}^-$, where $q_x^- = -q_x^+$, see the solid arrows in Fig.~\ref{fig:agnr_proc2}. After summation of all such
processes, the FT-LDOS acquires the left-right symmetry.

\begin{figure}
\includegraphics[width=0.49\columnwidth]{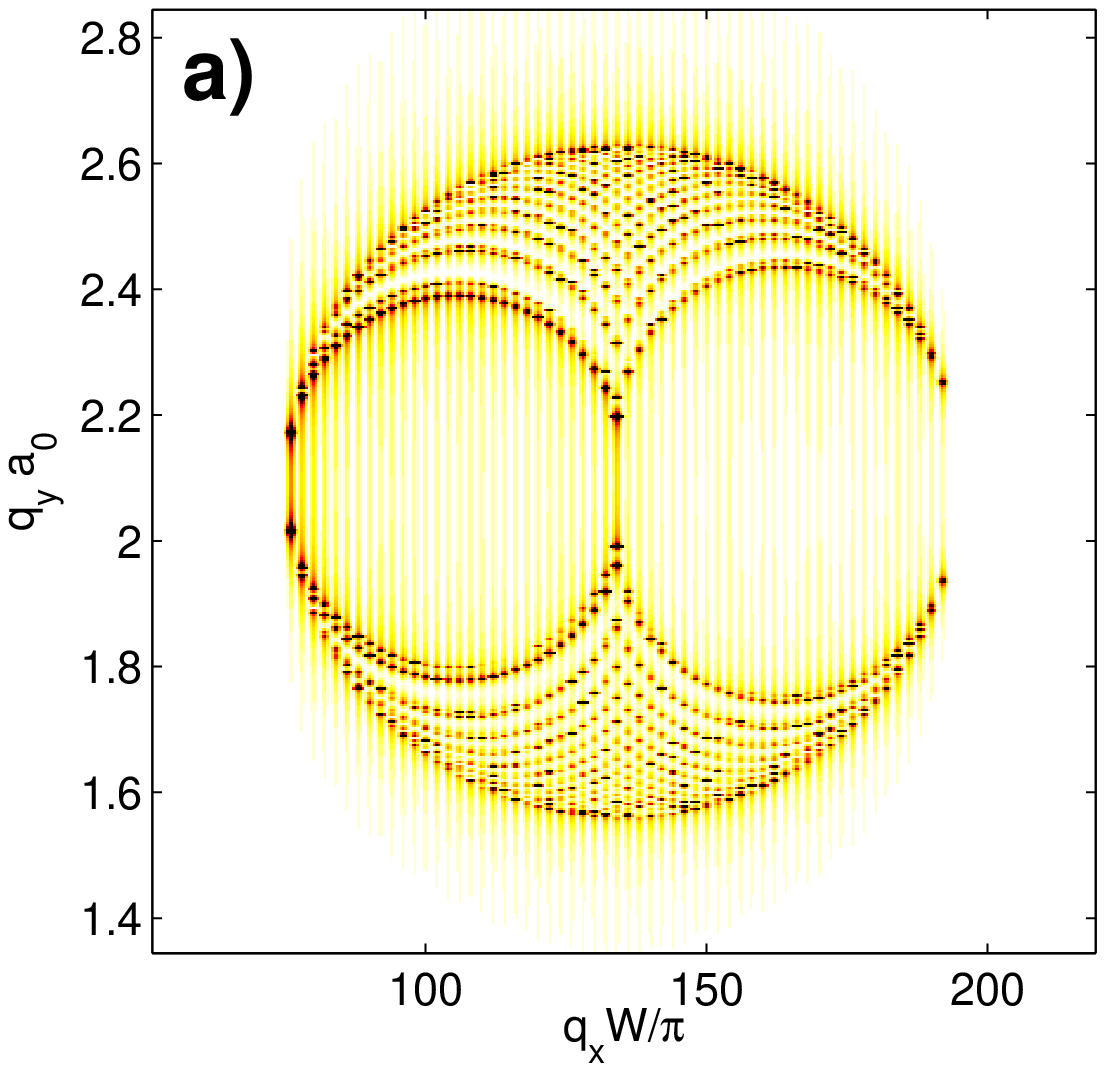}
\includegraphics[width=0.49\columnwidth]{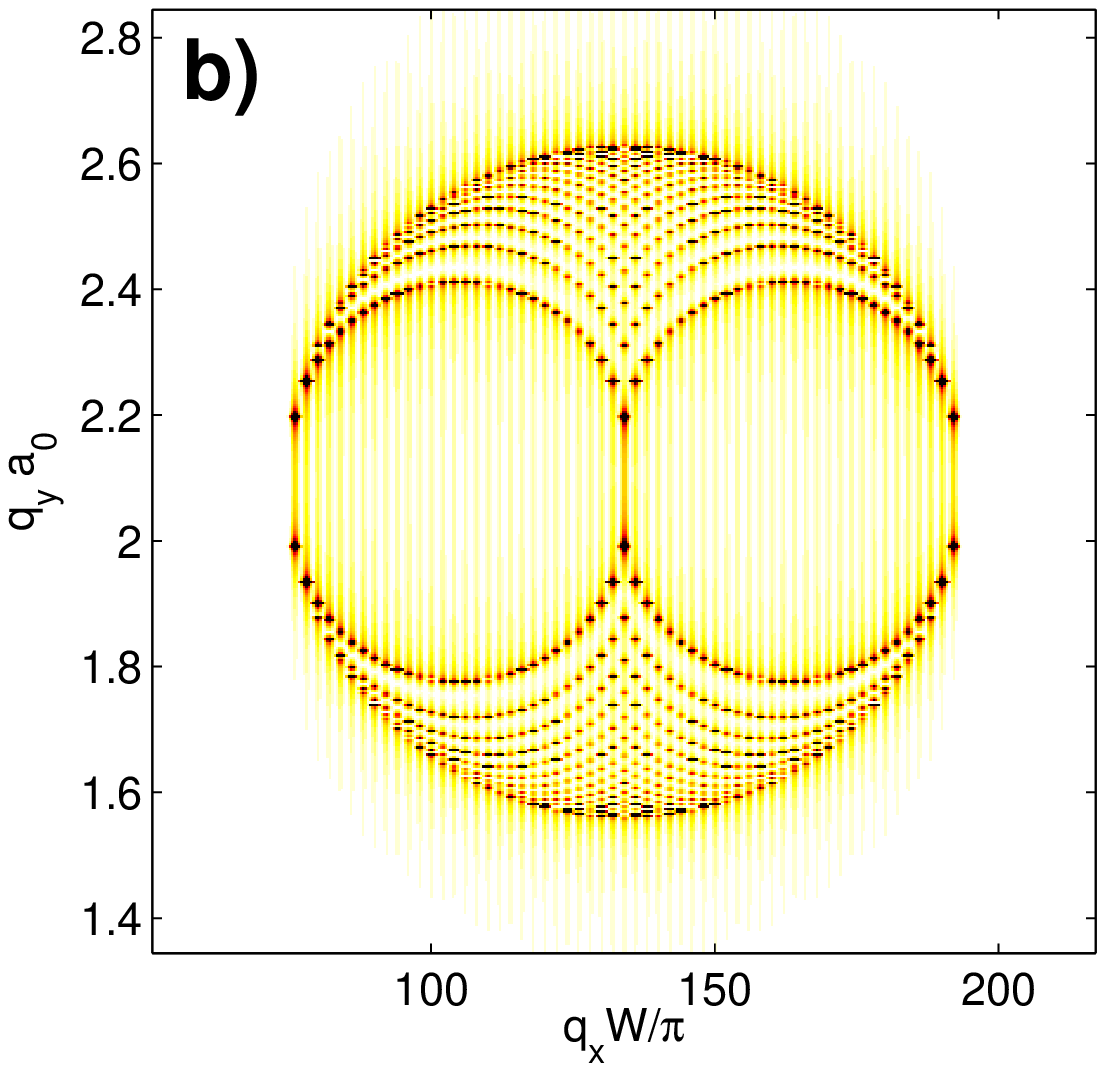}
\caption{Zoom-ins of the north-eastern circular feature in Fig.~\ref{fig:ftldos_full}. In (a), the AGNR is semiconducting and the 
left- and right-hand side is not mirror-symmetric. In (b), the ribbon is made metallic by removing 4 rows of carbon atoms, which 
restores the left-right symmetry again.}
\label{fig:ftldos_topright}
\end{figure}

As a consequence, the feature centered around $\vec{q}=0$ is always mirror symmetric by the above argument .
On the other hand, there is not necessarily a mirror symmetry within the other circular features (i.e. mirror symmetry with
respect to the individual cone centers).
For metallic AGNRs, the transverse modes are constructed from wavevectors symmetrically positioned with respect to 
the cone center (plus the metallic mode at the cone center).
See, for instance, the two wavefectors $k_{dn}$ and $k_{dn'}=-k_{dn}$ in Fig.~\ref{fig:agnr_proc2}.
For semiconducting AGNRs, the wavevectors are not
symmetrically positioned with respect to the cone center, i.e. $k_{dn'}\neq-k_{dn}$ for any $n'$.
Therefor, the inner structure of the circular features centered at finite $\vec{q}$ are symmetric
for metallic AGNRs and asymmetric for semiconducting AGNRs.
This is illustrated in Fig.~\ref{fig:ftldos_topright} for the semiconducting and metallic cases in (a) and (b), respectively.
We conclude that by looking at what symmetries there are in the FT-LDOS, one can extract information 
about whether or not an AGNR is metallic or not.

%
%

\section{Numerical simulations of the tight-binding model}
\label{Ch_numerics}

For our numerical simulations, we use a  tight-binding model described by the Hamiltonian
\begin{equation}
H = \sum_{i=1}^N \epsilon_i c^\dagger_i c_i + \sum_{i\neq j}^N \tau_{ij}c^\dagger_i c_j,
\end{equation}
where $c^\dagger_i$ and $c_i$  are creation and destruction operators for site $i$.
The onsite energy of site $i$ is denoted $\epsilon_i$, and the hopping amplitude between 
sites $j$ and $i$ is denoted $\tau_{ij}$. The number of atoms in the system is denoted $N$.
We assume that $\tau_{ij}$ is always zero except when the 
sites $i$ and $j$ are nearest neighbours.

The retarded Green's function matrix is defined as
\begin{equation}
\mathbf{G}(E) = \left[(E + i\eta)\mathbf{1} - \mathbf{H}\right]^{-1},
\end{equation}
where $\eta$ is a small positive number. Even though the Hamiltonian is sparse, when written down as a matrix
in site index space, direct inversion 
is not a viable alternative when the number of atoms $N$ grows large.
Instead of direct matrix inversion, we use our own implementation of a recent algorithm\cite{Kazymyrenko:2008hk} 
in which the system atoms are added one-by-one, in a recursive manner. This allows us 
to save both memory and time, and once we have found all the retarded propagators between the 
system leads and atom $i$ we can calculate the lesser Green's function, defined as
\begin{equation}
G^<_{ii}(E) = \sum_l f_l(E) \sum_{\alpha_l \beta_l} G_{i\alpha_l}(E)
\left[\Sigma^\dagger_l(E) - \Sigma_l(E)\right]_{\alpha_l \beta_l}
G^\dagger_{\beta_l i}(E),
\end{equation}
where $l$ is the lead number ($l = 1, 2$ in the case of a simple ribbon), and $\alpha_l$ and $\beta_l$
are indices running over all atoms belonging to the surface of lead $l$. Here, $f_l(E)$ and $\Sigma_l(E)$, are 
the distrubution function and the self-energy of lead $l$, respectively.

The local density of states on atom $i$ is found from
\begin{equation}
\rho_{i}(E) = -\frac{1}{\pi}\text{Im}G^<_{ii}(E),
\end{equation}
and the FT-LDOS is given by doing a discrete Fourier transform over all system atoms,
\begin{equation}
\mathcal{N}(\vec{q};E) = \frac{1}{N} \sum_{i=1}^N e^{-i \vec{r}_i \cdot \vec{q}} \rho_{i}(E),
\end{equation}
where $\vec{r}_i$ is the real space coordinate vector of atom $i$.

\begin{figure}
\includegraphics[width=0.4\columnwidth]{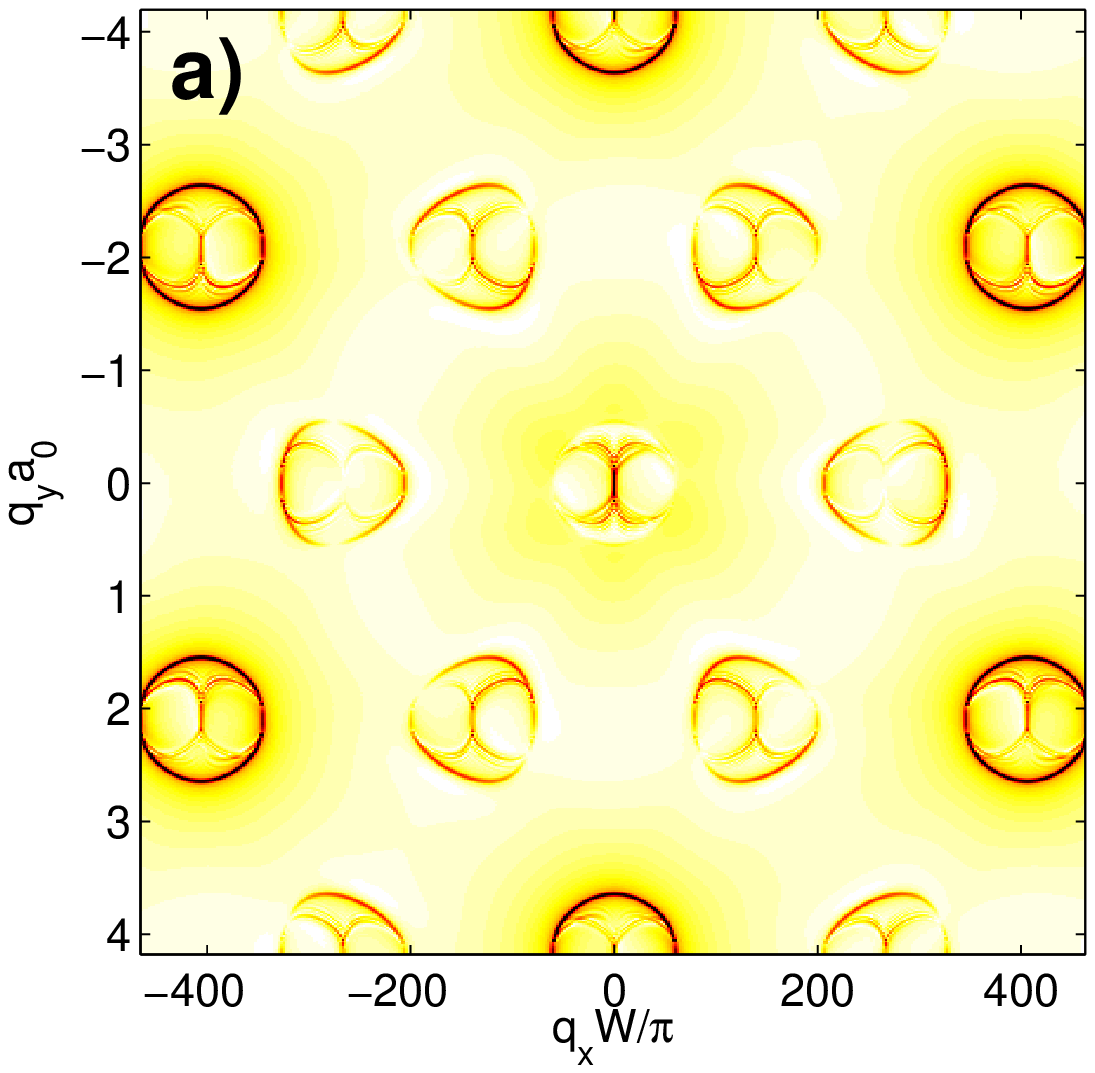}
\includegraphics[width=0.4\columnwidth]{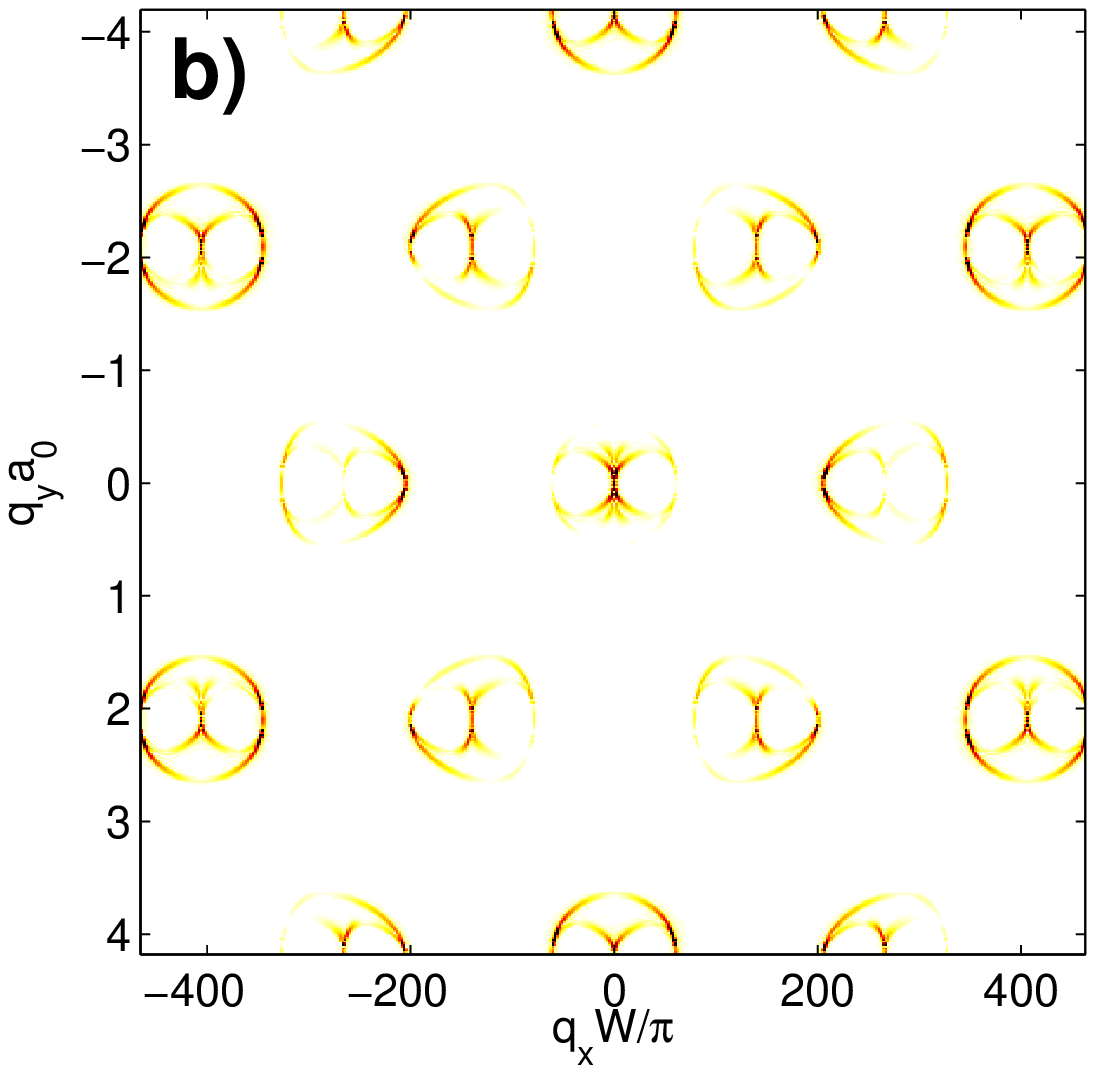}
\includegraphics[width=0.4\columnwidth]{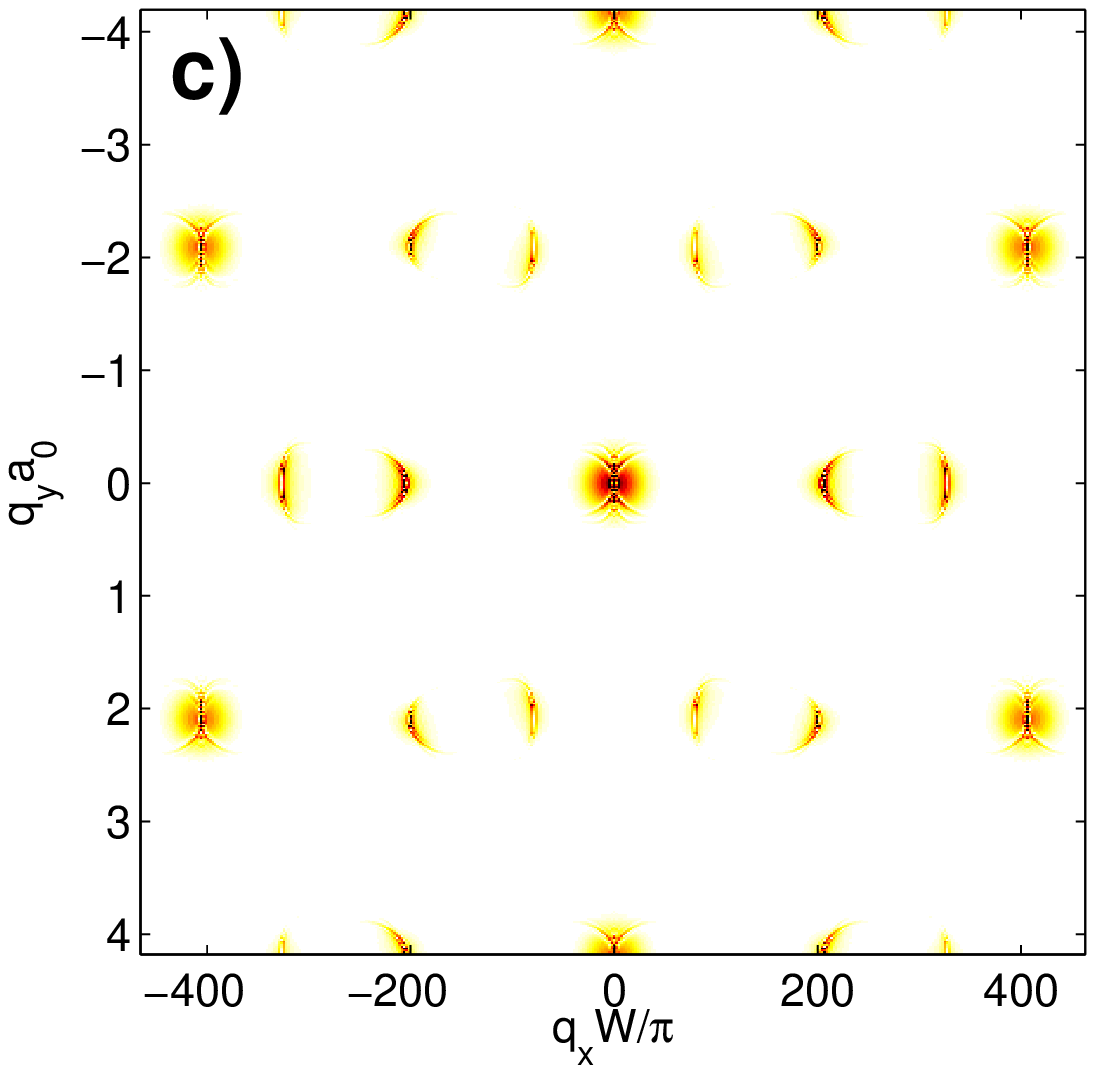}
\caption{Numerical tight-binding FT-LDOS of three AGNRs' ($N = 810$ atoms in the unit cell) with different impurity configurations. 
(a) Single impurity, (b) edge impurity, (c) smooth (Gaussian long-range) impurity, where $W \approx 50$ nm, $\gamma = 10|\tau|$, 
$E = 0.4|\tau|$ and 60 propagating channels.}
\label{fig:ftldos_tb}
\end{figure}

In Fig.~\ref{fig:ftldos_tb}(a), the result of such a tight-binding simulation is shown for a ribbon
and setup matching the one used in Fig.~\ref{fig:ftldos_full}, with a delta-like impurity placed in the middle 
($W \approx 50$ nm, 60 propagating channels and $x_i = W/2$). Upon inspection, we notice that the general features are 
similar compared with our analytical results. Some points, such as the outline of the central circle, are attenuated. The 
tight-binding ribbon do, however, show clear signs of trigonal warping due to the dispersion not being perfectly linear. 
In Fig.~\ref{fig:ftldos_tb}(b), we have moved the impurity to the edge of the ribbon and we notice that the resulting FT-LDOS image 
is not very different from the one with the impurity in the middle of the ribbon. In Fig.~\ref{fig:ftldos_tb}(c), we have made the 
impurity more gaussian shaped (long-range), which leads to suppressed scattering and attenuated features.
For bulk graphene, it is well known that a long range impurity can not scatter between valleys. In the FT-LDOS, 
the features centered at $\vec{q}=\vec{K}_d^{pm}$ are then absent. This is not the case here, since the armchair nanoribbon
has only one cone in its band structure.\cite{Wakabayashi:2009dh}

\begin{figure}
\includegraphics[width=0.4\columnwidth]{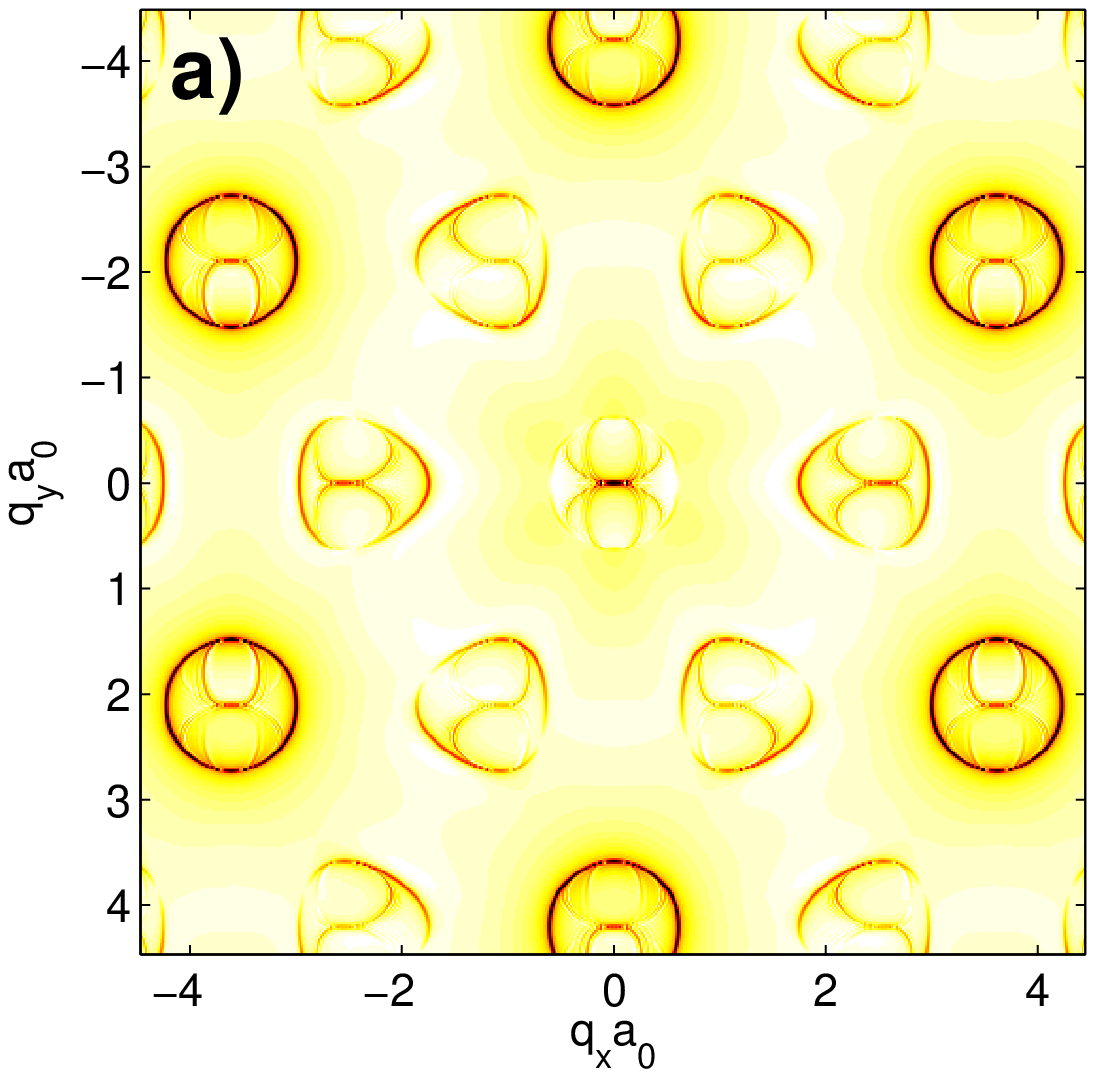}
\includegraphics[width=0.4\columnwidth]{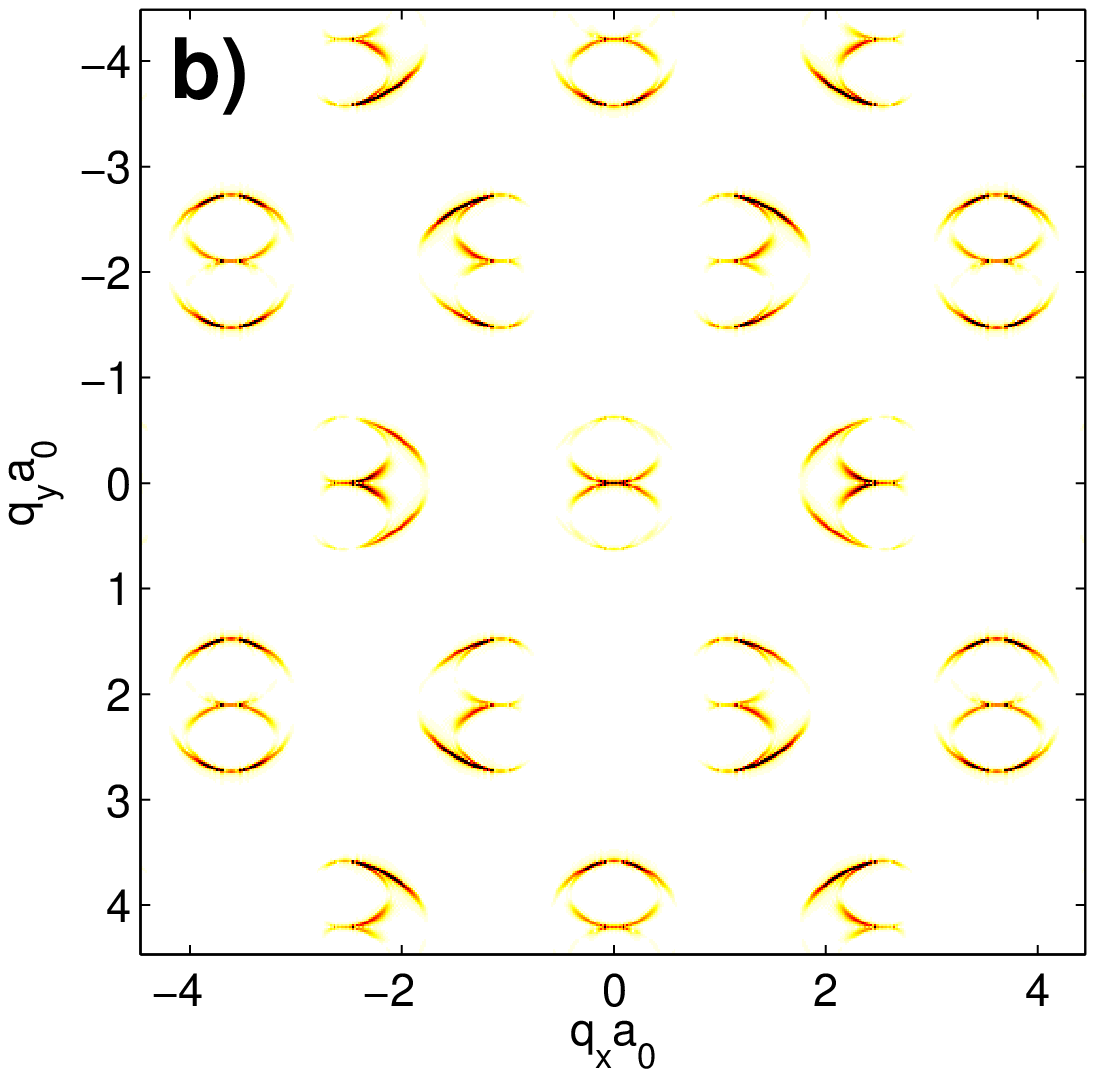}
\includegraphics[width=0.4\columnwidth]{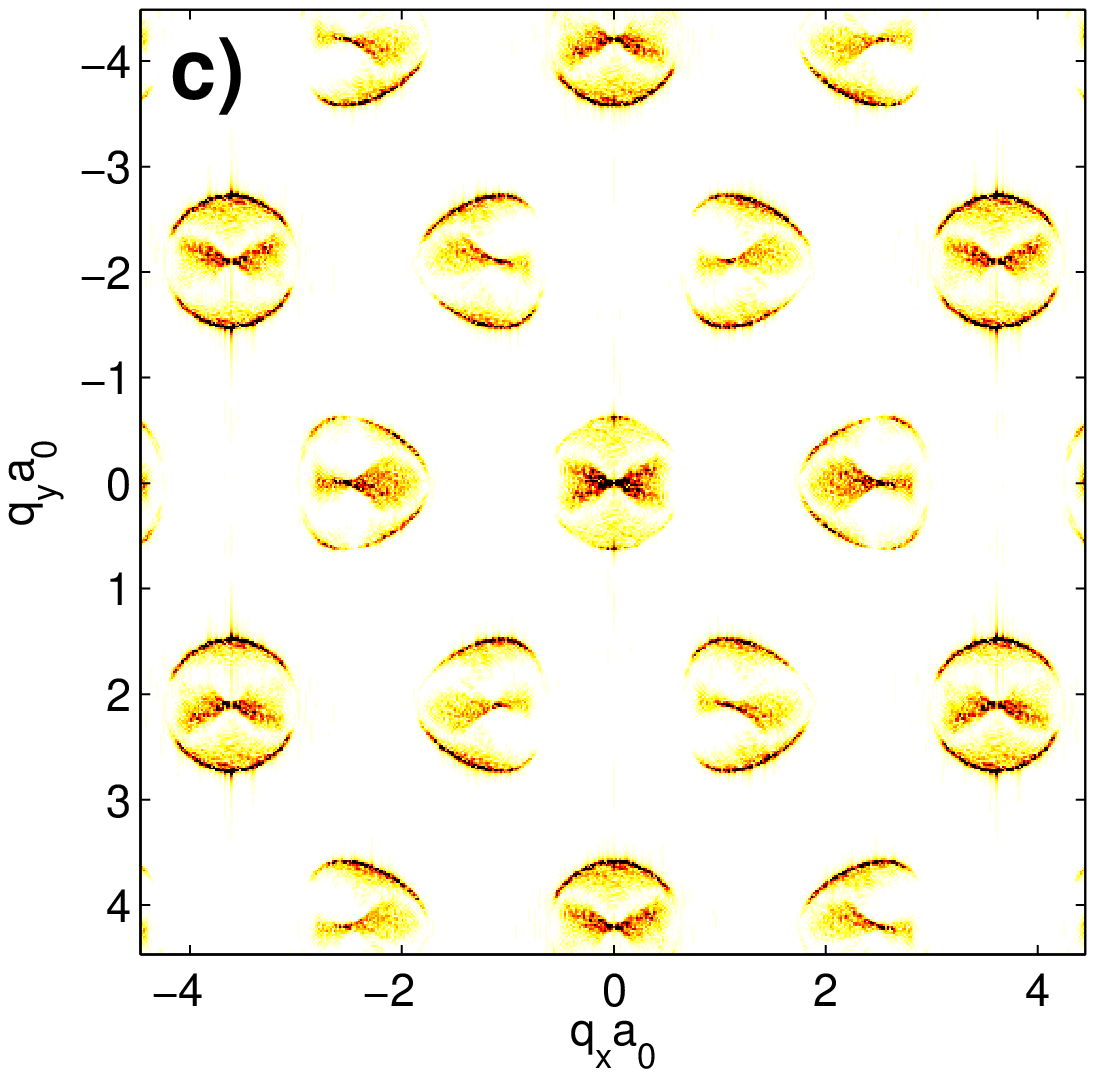}
\includegraphics[width=0.4\columnwidth]{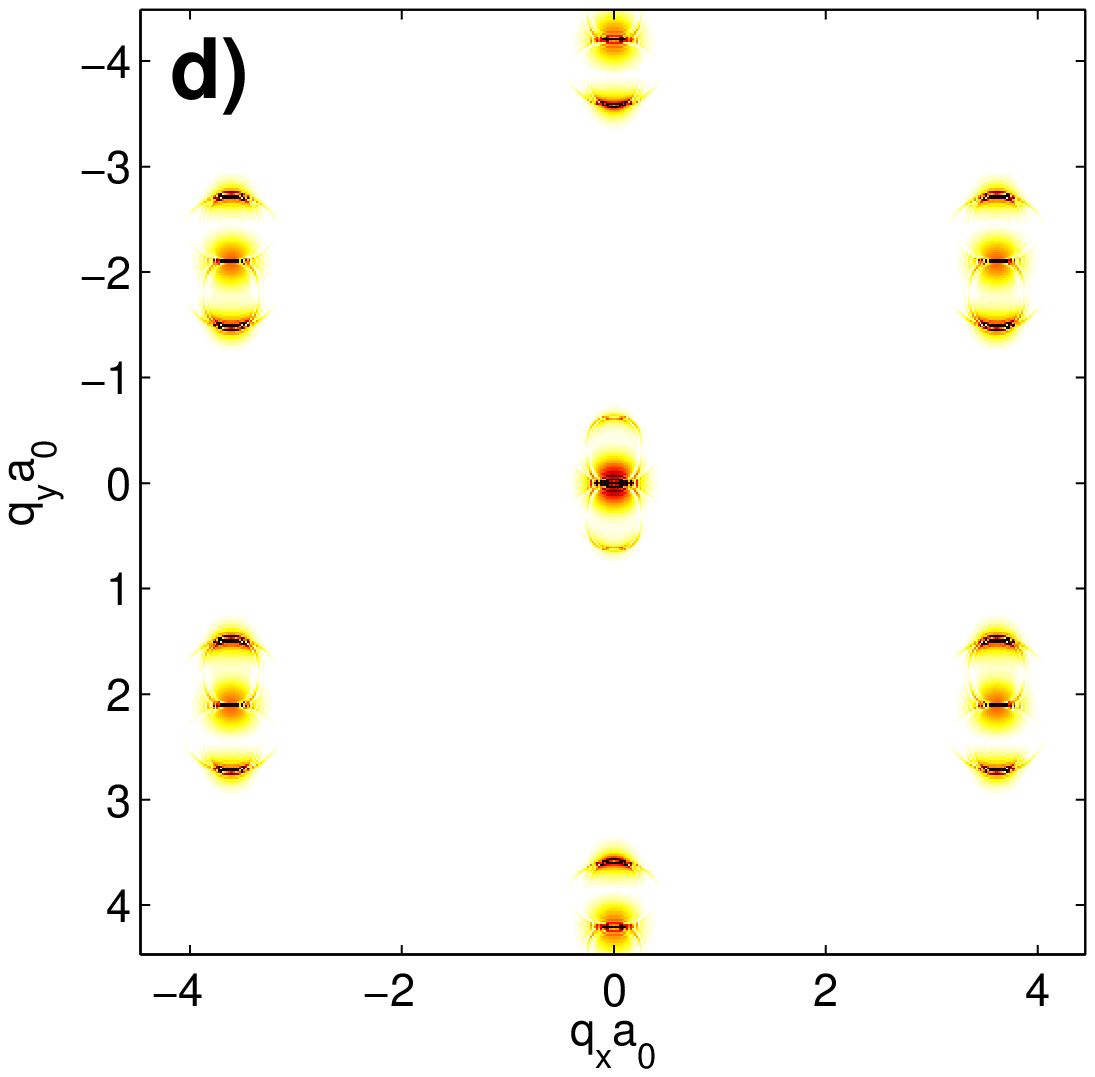}
\caption{Numerical tight-binding FT-LDOS of four ZGNRs' ($N = 468$ atoms in the unit cell) with different impurity configurations. 
(a) Single impurity, (b) edge impurity, (c) rough edges, (d) smooth impurity (Gaussian long-range), where $W \approx 50$ nm, 
$E = 0.45|\tau|$ $\gamma = 10|\tau|$, and $35$ propagating channels.}
\label{fig:ftldos_tbz}
\end{figure}

In Fig.~\ref{fig:ftldos_tbz}, we present results for the FT-LDOS of zigzag graphene nanoribbons (ZGNRs).
In this simulation the ribbon has $N = 468$ atoms in its unit cell ($W \approx 50$ nm),
$\gamma = 10\tau$ and $E = 0.45|\tau|$. This gives $35$ propagating modes. In Fig.~\ref{fig:ftldos_tbz}(a), the impurity is located 
in the middle of the ribbon and we see a pattern very similar to that of the same impurity configuration in an armchair ribbon, but 
with all features rotated $90$ degrees due to the different ribbon alignment (for ZGNRs', $k_y$ is quantized instead).
The result of a single impurity on the edge is shown in Fig.~\ref{fig:ftldos_tbz}(b), and in Fig.~\ref{fig:ftldos_tbz}(c) we show the 
spectra for a ribbon also having rough edges. In the last figure, Fig.~\ref{fig:ftldos_tbz} (d), we have used a gaussian shaped 
(long-range) impurity, and we here see clearly that inter-valley scattering is now fully supressed. Indeed, since the ZGNR has two
cones in its bandstructure, this case is similar to bulk graphene.

%
%

\section{Summary}
\label{Ch_summary}

In summary, we have presented results for the FT-LDOS of graphene nanoribbons with local scattering centers.
The interplay between size quantization and scattering leads to characteristic peaks that can be related to
the transverse modes of the nanoribbon. The main features include ring-like structures, analogous to the
case of an infinite 2D graphene sheet with a single scattering center. Inside the ring-like structure, new peaks 
appear that are related to inter and intra band scattering in the ribbon. 
We have presented analytic results for the electron propagator in armchair nanoribbons in the Dirac
  approximation, including a single scattering center within a
  T-matrix formulation. We have also extended the
  investigation with numerics obtained with an atomistic recursive
  Green's function approach. The spectral signatures of the atomistic
  approach include the lifting of degeneracies of transverse modes in
  the Dirac approximation, as well as effects of trigonal warping.
  The impurity induced oscillations in the local density of states are
  not decaying at large distance in few-mode nanoribbons.

%
%
\section{Acknowledgements}

This work has been supported by SSF, the Swedish Foundation for Strategic Research,
and the EU through the FP7 project ConceptGraphene.\\

\newpage
\appendix

%
%

\section{Nanoribbon in a 2DEG}
\label{Appendix_2DEG}

\subsection{Unperturbed Green's function}
For a 2DEG confined in the x-direction, creating a ribbon of width $W$, the wave 
functions can be written as
\begin{equation}
\phi_n(\rv) = e^{ik_yy}\chi_n(x),
\end{equation}
where $\rv = (x,y)$ and $n$ is the mode number associated with the transverse 
eigenfunctions (assuming infinitely high confining walls at $x=0$ and $x=W$) given by
\begin{equation}
\chi_n(x) = \sqrt{\frac{2}{W}}\sin(k_nx),
\end{equation}
with the corresponding eigenenergies
\begin{equation}
\epsilon_n(k_y) = \frac{\hbar^2}{2\mu}(k_n^2 + k_y^2).
\end{equation}
Here, $k_n = n\pi/W$ is the transverse momentum, $k_y$ the longitudinal momentum and 
$\mu$ the electron mass. Using these wave functions we may construct the free propagator, 
or Green's function, of an electron (having energy $E^+ = E + i\eta$, in the limit 
$\eta \rightarrow 0^+$) between the points $\rvp$ and $\rv$, in mode $n$, as
\begin{equation}
\begin{split}
g_n(\rv,\rvp;E) &= 
\int_{-\infty}^\infty \frac{dk_y}{2\pi} \frac{\phi_n(\rv)\phi_n^*(\rvp)}{E^+ - \epsilon_n(k_y)} \\
&= \chi_n(x)\chi_n(x^\prime) \Gamma_n(y,y^\prime;E),
\end{split}
\end{equation}
where
\begin{equation}
\label{eq:gamma1}
\Gamma_n(y,y^\prime;E) = \int_{-\infty}^\infty \frac{dk_y}{2\pi} \frac{e^{ik_y(y-y^\prime)}}{E^+ - \frac{\hbar^2}{2\mu}(k_y^2 + k_n^2)}.
\end{equation}
This integral can be evaluated using standard contour integration techniques,\cite{Eco} giving us that
\begin{equation}
\Gamma_n(y,y^\prime;E) = 
\begin{cases}
-i\frac{\mu}{\hbar^2}\frac{e^{i\kappa_n(E)|y-y^\prime|}}{\kappa_n(E)}  & \text{if} \quad E > E_n, \\
-\frac{\mu}{\hbar^2}\frac{e^{-\kappa_n(E)|y-y^\prime|}}{\kappa_n(E)} & \text{if} \quad E < E_n,
\end{cases}
\end{equation}
and
\begin{equation}
\kappa_n(E) = \sqrt{\frac{2\mu}{\hbar^2}|E - E_n|},
\end{equation}
where $E_n = (\hbar^2/2\mu)k_n^2$.

\subsection{Green's function, one impurity}
We introduce a single impurity modelled by an impurity potential with
matrix elements $V_{nm}(\rv)$. The perturbed propagator for an
electron going from position $\rvp$ to $\rv$, while changing mode from
$m$ to $n$, can then be written using the Dyson equation as\cite{Bagwell:2012te}
\begin{equation}
\begin{split}
G_{nm}(\rv,\rvp;E) &= g_n(\rv,\rvp;E)\delta_{nm} \\
& + \underbrace{\sum_l g_n(\rv,\rv_i;E) V_{nl}(\rv_i) G_{lm}(\rv_i,\rvp;E)}_{=\tilde{G}_{nm}(\rv,\rvp;E)} \\
& = g_n(\rv,\rvp;E)\delta_{nm} + \tilde{G}_{nm}(\rv,\rvp;E).
\end{split}
\end{equation}
Here we assume that the impurity is positioned at $\rv_i = (x_i,y_i)$ and that its potential 
is highly localized (delta-function shaped) so that all matrix elements but $V_{nm}(\rv_i)$ 
are zero. After introducing
\begin{equation}
T_{nm}(\rv_i;E) = V_{nm}(\rv_i) + \sum_l V_{nl}(\rv_i) g_l(\rv_i,\rv_i;E) T_{lm}(\rv_i;E),
\end{equation}
the scattering part of the Dyson equation can be rewritten on the T-matrix form
\begin{equation}
\label{eq:2DEGT}
\tilde{G}_{nm}(\rv,\rvp;E) = g_n(\rv,\rv_i;E)T_{nm}(\rv_i;E)g_m(\rv_i,\rvp;E),
\end{equation}
where
\begin{equation}
T_{nm}(\rv_i;E) =  V_{nm}(\rv_i) + \sum_lV_{nl}(\rv_i)g_l(\rv_i,\rv_i;E)T_{lm}(\rv_i;E).
\end{equation}
Since the impurity is highly localized in position space, we may further assume that it will
scatter equally between all different modes $n$ and $m$ and we have that 
$V_{nm}(\rv_i) = V(\rv_i) = \gamma$ where $\gamma$ is the impurity strength. Using this 
assumption, it  follows that $T_{nm}(\rv;E) = T(\rv;E)$ and we find that
\begin{equation}
\label{eqn:T_2deg}
\begin{split}
T(\rv_i;E) &= V(\rv_i) + V(\rv_i)\left[\sum_lg_l(\rv_i,\rv_i;E)\right]T(\rv_i;E) \\
           &= \frac{\gamma}{1 - \gamma \sum_l g_l(\rv_i,\rv_i;E)} \\
           &= \frac{1}{1/\gamma + \sigma_e(E) + i\sigma_p(E)},
\end{split}
\end{equation}
where
\begin{equation}
\label{eqn:sigmas_2deg}
\sigma_{e/p}(E) = \frac{\mu}{\hbar^2} \sum_{l\in e/p} \frac{\chi^2_l(x_i)}{\kappa_l(E)}
\end{equation}
and $e$ and $p$ are the sets of all evanescent ($E < E_l$) and propagating 
($E > E_l$) modes. Inserting the above expression for $T(\rv_i;E)$ back 
into Eq. (\ref{eq:2DEGT}) allows us to solve for $\tilde{G}_{nm}(\rv,\rvp;E)$ 
and consequently for $G_{nm}(\rv,\rvp;E)$.
\subsection{Fourier transformed density of states}
Once the perturbed propagator is known, the change in the local density of states (LDOS) due 
to scattering is given by
\begin{equation}
\begin{split}
\tilde{\rho}(\rv;E) &= -\frac{1}{\pi}\sum_{nm}\Imt\left[\tilde{G}_{nm}(\rv,\rv;E)\right] \\
&= -\frac{1}{\pi}\sum_{nm} \mathcal{K}_{nm}(E)\tilde{\rho}^x_{nm}(x) \tilde{\rho}^y_{nm}(y;E),
\end{split}
\end{equation}
where
\begin{equation}
\mathcal{K}_{nm}(E) = \left(\frac{\mu}{\hbar^2}\right)^2 \frac{1}{(1 + \sigma_e(E))^2 + \sigma_p^2(E)} \frac{\chi_n(x_i)\chi_m(x_i)}{\kappa_n(E)\kappa_m(E)},
\end{equation}
\begin{equation}
\tilde{\rho}^x_{nm}(x) = \chi_n(x)\chi_m(x)
\end{equation}
and
\begin{equation}
\tilde{\rho}^y_{nm}(y;E) =
\begin{cases}
-f_{sc}(\kappa_n(E),\kappa_m(E)) & \text{if $n,m \in p$}, \\
f_{cs}(\kappa_n(E),0)e^{-\kappa_m(E)|y-y_i|} & \text{if $n\in p, m\in e$}, \\
f_{cs}(0,\kappa_m(E))e^{-\kappa_n(E)|y-y_i|} & \text{if $n\in e, m\in p$}, \\
-\sigma_p(E) e^{-(\kappa_n(E) + \kappa_m(E))|y-y_i|} & \text{if $n,m \in e$},
\end{cases}
\end{equation}
where
\begin{equation}
\begin{split}
f_{cs}(\kappa_1,\kappa_2) &= (1 + \sigma_e)\cos\left[(\kappa_1+\kappa_2)|y-y_i|\right] \\
													&+ \sigma_p\sin\left[(\kappa_1+\kappa_2)|y-y_i|\right],
\end{split}
\end{equation}
and
\begin{equation}
\begin{split}
f_{sc}(\kappa_1,\kappa_2) &= 
(1 + \sigma_e)\sin\left[(\kappa_1+\kappa_2)|y-y_i|\right] \\
&- \sigma_p\cos\left[(\kappa_1+\kappa_2)|y-y_i|\right].
\end{split}													
\end{equation}
\newline
When taking the Fourier transform of the scattering LDOS, we want to 
be able to resolve differences in $x$-momenta equal to or greater than 
$\pi/W$ (since this is the separation in $k_x$, or $k_n$, between two adjacent 
subbands). This requires us to integrate over the interval $[-W,W]$ and 
we extend the function $\tilde{\rho}^x_{nm}(x)$ such that it is even 
with respect to the origin. The Fourier transform is then defined as
\begin{equation}
\begin{split}
\Nt_{nm}(\vec{q};E) &= \mathcal{K}_{nm}(E) \times \\
& \times \sum_{n^\prime = -\infty}^\infty \delta\left(\frac{q_x}{\pi} - \frac{n^\prime}{W}\right) \int_{-W}^W \frac{dx}{2W} e^{-iq_x x}\tilde{\rho}^x_{nm}(x)\times \\
& \times \int_{-\infty}^\infty \frac{dy}{2\pi}e^{-iq_y y}\tilde{\rho}^y_{nm}(y;E),
\end{split}
\label{eq:fourier}
\end{equation}
where the comb function fixes $q_x$ to multiples of $\pi/W$. The x-part of the Fourier integral is
\begin{equation}
\begin{split}
\int_{-W}^W \frac{dx}{2W} e^{-iq_x x} \tilde{\rho}^x_{nm}(x) &= \frac{1}{2W}\left(\delta_{l,-n-m} + \delta_{l,n+m} \right. \\
&\quad \left. - \delta_{l,-n+m} - \delta_{l,n-m}\right),
\end{split}
\end{equation}
independent of if $n$ and $m$ are evanescent or propagating modes. 
\newline
\newline
The y-part will depend on 
mode types. We have already shown what happens when $n,m \in p$. In addition, if $n,m \in e$ we get that
\begin{equation}
\int_{-\infty}^\infty \frac{dy}{2\pi} e^{-i q_y y} \tilde{\rho}^y_{nm}(y;E) = -e^{-iq_yy_i}\frac{\sigma_p}{\pi}\frac{\kappa_n(E) + \kappa_m(E)}{q^2_y + (\kappa_n(E)+\kappa_m(E))^2}.
\end{equation}
If $n\in p, m \in e$ then
\begin{equation}
\int_{-\infty}^\infty \frac{dy}{2\pi} e^{-i q_y y} \tilde{\rho}^y_{nm}(y;E) = \frac{e^{-iq_yy_i}}{2\pi}\left[S^y_{pe}(\kappa_n(E)-q_y,\kappa_m(E)) + S^y_{pe}(\kappa_n(E)+q_y,\kappa_m(E))\right],
\end{equation}
where
\begin{equation}
S^y_{pe}(a,b) = \frac{(1/\gamma + \sigma_e(E))b + \sigma_p(E)a}{b^2 + a^2}.
\end{equation}
If $n\in e, m\in p$ we just need to interchange the $n$ and $m$ in the expression above.
\begin{equation}
\end{equation}
\color{black}
%
%

\section{Armchair graphene nanoribbon}
\label{Appendix_armchair}
\begin{figure}[t]
\includegraphics[width=0.49\columnwidth]{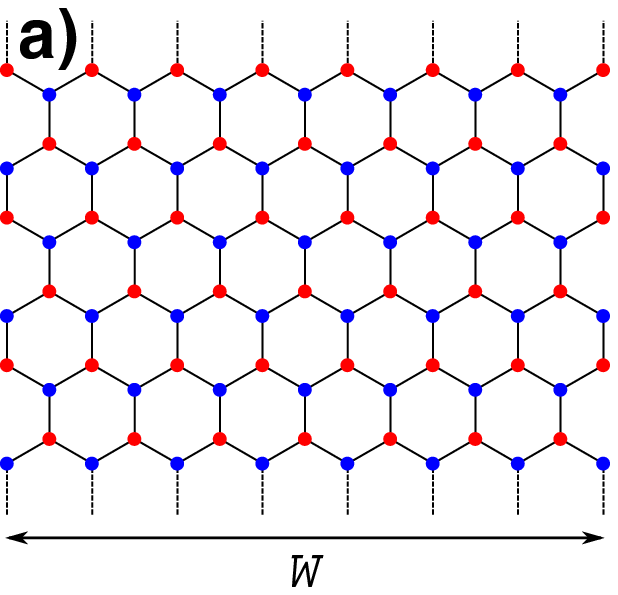}
\includegraphics[width=0.49\columnwidth]{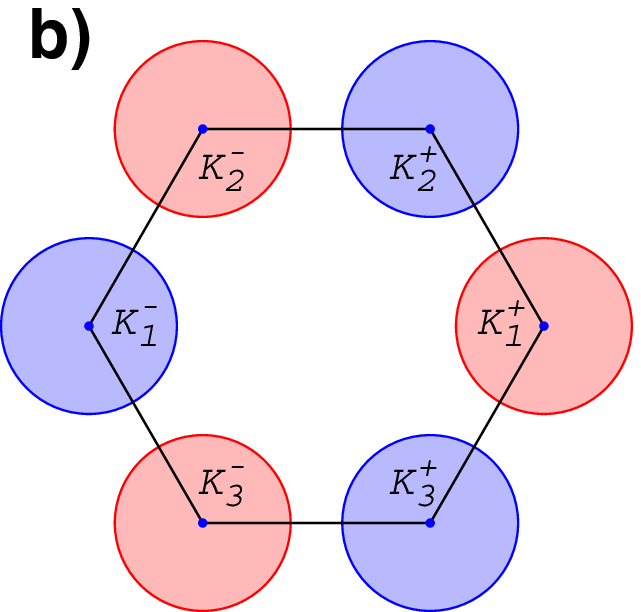}
\caption{(a) The geometry of the armchair nanoribbon. (b) The first
  Brillouin zone with three sets of Dirac cones.}
\label{AC_fig}
\end{figure}

In this appendix we first derive an analytic expression for the
Green's function of an armchair nanoribbon with a single impurity. For
the geometry, see Fig.~\ref{AC_fig}(a). We then derive the Fourier
transformed density of states.

\subsection{Unperturbed Green's function}
The first Brillouin zone (1BZ) of graphene contains one pair of
inequivalent Dirac cones. It is necessary, however, to include three
pairs of cones [see Fig.~\ref{AC_fig}(b)] in order to incorporate all
scattering events. The cones are located at
$\vec{K}_1^{\pm} = (\pm K_x,0)$, $\vec{K}_2^{\pm} = (\pm K_x/2,K_y)$ and $\vec{K}_3^{\pm} = (\pm K_x/2,-K_y)$,
where $K_x = 4\pi/3a$ and $K_y = 2\pi/3a_0$. The distance between two neighbouring atoms is denoted $a_0$, while
the lattice constant is denoted $a = \sqrt{3}a_0$. This gives us three
sets ($d = 1,2,3$) of wave function spinors,\cite{Brey:2006cb}
\begin{equation}
\vec{\Psi}_{dn}(\rv) = \begin{pmatrix} \Psi^A_{dn}(y) \\ \Psi^B_{dn}(y) \end{pmatrix} \chi_{dn}(x),
\end{equation}
where the longitudinal wave function components are
\begin{equation}
\Psi^A_{dn}(y) = \lambda \frac{(k_y + ik_{dn})}{\sqrt{k_y^2 + k_{dn}^2}}e^{i(K_{dy} + k_y)y},
\end{equation}
\begin{equation}
\Psi^B_{dn}(y) = i e^{i(K_{dy} + k_y)y},
\end{equation}
and the transverse wave function is
\begin{equation}
\chi_{dn}(x) = 2C\sin\left[(K_{dx} + k_{dn})x\right].
\end{equation}
The integer number $n$ labels the quantized transverse momentum $k_{dn}
= n\pi/W - K_{dx}$ in cone pair $d$. For each mode $n$, we have
positive and negative energy subbands $\epsilon_{dn\lambda}(k_y) =
\lambda v_f\sqrt{k_y^2 + (k_{dn})^2}$ labeled by $\lambda=\pm 1$. The
Fermi velocity is $v_f = 3a_0|t|/2$, where $t$ is the nearest neighbor
tight-binding hopping energy. The wave functions have been normalized
through a normalization constant $C = \sqrt{1/4W}$ found from the
condition $\int_0^W dx |\chi_{dn}(x)|^2 = 1/2$. Thus, $\chi_{dn}(x) = \chi_n(x) = \sqrt{1/W}\sin(n\pi/Wx)$.
\newline
The free propagator for band $n$ (in cone pair $d$) is computed as
\begin{equation}
\begin{split}
\mathbf{g}_{dn}(\rv,\rvp) &=
\sum_{\lambda = \pm 1}
\int_{-\infty}^\infty \frac{dk_y}{2\pi}
\frac{\vec{\Psi}_{dn}(\rv)\vec{\Psi}_{dn}^\dagger(\rvp)}{E^+ - \epsilon_{dn\lambda}(k_y)} \\
&= \chi_{n}(x)\chi_{n}(x^\prime)
\begin{pmatrix}
\Gamma_{dn}^{AA}(y,y^\prime;E) & \Gamma_{dn}^{AB}(y,y^\prime;E) \\
\Gamma_{dn}^{BA}(y,y^\prime;E) & \Gamma_{dn}^{BB}(y,y^\prime;E)
\end{pmatrix},
\end{split}
\end{equation}
where
\begin{equation}
\Gamma_{dn}^{AA/BB}(y,y^\prime;E) = 2E e^{iK_{dy}(y-y^\prime)}\int_{-\infty}^\infty \frac{dk_y}{2\pi} \frac{e^{ik_y(y-y^\prime)}}{(E^+)^2 - v_f^2(k_y^2 + k_{dn}^2)},
\end{equation}
and
\begin{equation}
\Gamma_{dn}^{AB/BA}(y,y^\prime;E) = \mp 2iv_f e^{iK_{dy}(y-y^\prime)} \int_{-\infty}^\infty \frac{dk_y}{2\pi} \frac{(k_y \pm ik_{dn})e^{ik_y(y-y^\prime)}}{(E^+)^2 - v_f^2(k_y^2 + k_{dn}^2)}.
\end{equation}
After contour integration, we find the final form to be
\begin{equation}
\Gamma_{dn}^{AA/BB}(y,y^\prime;E) = -i\frac{|E|}{v_f^2} e^{iK_{dy}(y-y^\prime)}\frac{e^{i\sgn{E}\kappa_{dn}(E)|y-y^\prime|}}{\kappa_{dn}(E)},
\end{equation}
and
\begin{equation}
\Gamma_{dn}^{AB/BA}(y,y^\prime;E) = -\frac{1}{v_f}e^{iK_{dy}(y-y^\prime)}\left[\frac{i\sgn{E}k_{dn}}{\kappa_{dn}(E)} \pm \sgn{y-y^\prime}\right]e^{i\sgn{E}\kappa_{dn}(E)|y-y^\prime|},
\end{equation}
where
\begin{equation}
\kappa_{dn} = \sqrt{|(E/v_f)^2 - k_{dn}^2|}.
\end{equation}
In the above formulas, we have assumed that $n$ is a propagating mode (e.g. $|E/v_f| > |k_{dn}$). If mode $n$ is evanescent ($|E/v_F| < |k_{dn}|$), we have to modify the longitudinal momentum so that $\kappa_{dn} \rightarrow i\text{sgn}(E)\kappa_{dn}$.
\subsection{Green's function, one impurity}

For the graphene armchair ribbon, we select an impurity fully
localized on the A-sublattice, scattering equally between all modes
$n$ and $m$. The matrix elements of the impurity potential then is
\begin{equation}
\label{eq:Vgr}
\mathbf{V}_{nm}(\rv_i,\rv_i) = \mathbf{V}(\rv_i,\rv_i) = \gamma \begin{pmatrix} 1 & 0 \\ 0 & 0 \end{pmatrix},
\end{equation}
where $\gamma$ is the impurity strength. The T-matrix equation is
written down in analogy to the 2DEG case, but for graphene it acquires
a 2x2 matrix structure. For the potential in Eq.~(\ref{eq:Vgr}), we
get
\begin{equation}
\begin{split}
\m{T}(\rv_i,\rv_i;E) &= \m{V}(\rv_i,\rv_i) + \m{V}(\rv_i,\rv_i)\left[\sum_d\sum_l \m{g}_{dl}(\rv_i,\rv_i;E) \right]\m{T}(\rv_i,\rv_i) \\
&= \frac{1}{1/\gamma + \sigma_e(E) + i\sigma_p(E)} \begin{pmatrix} 1 & 0 \\ 0 & 0 \end{pmatrix},
\end{split}
\end{equation}
where
\begin{equation}
\sigma_{p}(E) = \frac{|E|}{v_f^2}\sum_d\sum_{l\in p} \frac{\chi_{l}^2(x_i)}{\kappa_{dl}(E)}
\end{equation}
and
\begin{equation}
\sigma_{e}(E) = \frac{E}{v_f^2}\sum_d\sum_{l\in e} \frac{\chi_{l}^2(x_i)}{\kappa_{dl}(E)}.
\end{equation}

The letters $e$ and $p$ denotes sets of evanescent and propagating modes,
respectively. The Dyson equation for the Green's function can now be
written as
\begin{equation}
\m{G}_{dncm}(\rv,\rvp) = \m{g}_{dn}(\rv,\rvp)\delta_{nm} + \m{\tilde{G}}_{dncm}(\rv,\rvp)
\end{equation}
where
\begin{equation}
\m{\tilde{G}}_{dncm}(\rv,\rvp;E) = \m{g}_{dn}(\rv,\rv_i;E)\m{T}(\rv_i,\rv_i;E)\m{g}_{cm}(\rv_i,\rvp;E).
\end{equation}
The scattering part $\m{\tilde{G}}_{dncm}(\rv,\rvp;E)$ takes the form
\begin{equation}
\begin{split}
\tilde{\m{G}}_{dncm}(\rv,\rvp;E) &= \frac{1}{1/\gamma + \sigma_e(E) + i\sigma_p(E)}
\begin{pmatrix}
g^{AA}_{dn}(\rv,\rv_i;E)g^{AA}_{cm}(\rv_i,\rvp;E) & g^{AA}_{dn}(\rv,\rv_i;E)g^{AB}_{cm}(\rv_i,\rvp;E) \\
g^{BA}_{dn}(\rv,\rv_i;E)g^{AA}_{cm}(\rv_i,\rvp;E) & g^{BA}_{dn}(\rv,\rv_i;E)g^{AB}_{cm}(\rv_i,\rvp;E)
\end{pmatrix} \\
&= 
\frac{\chi_{n}(x)\chi_{n}(x_i)\chi_{m}(x_i)\chi_{m}(x^\prime)}{1/\gamma+\sigma_e(E) + i\sigma_p(E)} 
\begin{pmatrix}
\Gamma^{AA}_{dn}(y,y_i;E)\Gamma^{AA}_{cm}(y_i,y^\prime;E) & \Gamma^{AA}_{dn}(y,y_i;E)\Gamma^{AB}_{cm}(y_i,y^\prime;E) \\
\Gamma^{BA}_{dn}(y,y_i;E)\Gamma^{AA}_{cm}(y_i,y^\prime;E) & \Gamma^{BA}_{dn}(y,y_i;E)\Gamma^{AB}_{cm}(y_i,y^\prime;E)
\end{pmatrix}.
\end{split}
\end{equation}
For the computation of the local density of states, we need the two
diagonal components. Their explicit forms ($n,m \in p$) are
\begin{equation}
\begin{split}
\tilde{G}^{AA}_{dncm}(\rv,\rvp;E) &= 
-\frac{1}{1/\gamma + \sigma_e(E) + i\sigma_p(E)}\left(\frac{|E|}{v_f^2}\right)^2\chi_{n}(x)\chi_{n}(x_i)\chi_{m}(x_i)\chi_{m}(x^\prime)\times \\
& \times e^{iK_{dy}(y-y_i)}e^{iK_{cy}(y_i - y^\prime)}\frac{e^{i\sgn{E}(\kappa_{dn}(E)|y-y_i| +\kappa_{cm}(E)|y_i-y^\prime|)}}{\kappa_{dn}(E)\kappa_{cm}(E)}
\label{GAA}
\end{split}
\end{equation}
and
\begin{equation}
\begin{split}
\tilde{G}^{BB}_{dncm}(\rv,\rvp;E) &= 
\frac{1}{1/\gamma + \sigma_e(E) + i\sigma_p(E)}\left(\frac{1}{v_f}\right)^2 \chi_{n}(x)\chi_{n}(x_i)\chi_{m}(x_i)\chi_{m}(x^\prime)e^{iK_{dy}(y-y_i)}e^{iK_{cy}(y_i - y^\prime)}  \times \\
& \times e^{i\sgn{E}(\kappa_{dn}(E)|y-y_i| +\kappa_{cm}(E)|y_i-y^\prime|)} \left[\frac{i\sgn{E}k_{dn}}{\kappa_{dn}(E)} - \sgn{y-y_i}\right] \left[\frac{i\sgn{E}k_{cm}}{\kappa_{cm}(E)} + \sgn{y_i-y^\prime}\right].
\label{GBB}
\end{split}
\end{equation}

\subsection{Density of states}

The scattering correction to the local density of states can be
computed separately for the two sublattices, and is given by
\begin{equation}
\begin{split}
\tilde{\rho}_{A/B}(\rv;E) &= 
-\frac{1}{\pi}\sum_{dc}\sum_{nm}\Imt\left[\tilde{G}_{dncm}^{AA/BB}(\rv,\rv;E)\right] \\
&= -\frac{1}{\pi}\sum_{dc}\sum_{nm}\mathcal{K}^{A/B}_{dncm}(E)\tilde{\rho}^x_{nm}(x)\tilde{\rho}^{(A/B)y}_{dncm}(y;E),
\end{split}
\label{LDOS_graphene}
\end{equation}
where $\tilde{\rho}^x_{nm}(x) = \chi_{n}(x)\chi_{m}(x)$. The A/B sublattice
corrections are found by substituting Eq. (\ref{GAA}) and Eq. (\ref{GBB}) respectively in
Eq. (\ref{LDOS_graphene}). The results are very similar (the A correction being almost identical) to the 2DEG
case, and for $n,m \in p$ we find that
\begin{equation}
\mathcal{K}^A_{dncm}(E) = \frac{1}{(1/\gamma+\sigma_e(E))^2 + \sigma_p^2(E)}\left(\frac{|E|}{v_f^2}\right)^2\frac{\chi_{n}(x_i)\chi_{m}(x_i)}{\kappa_{dn}(E)\kappa_{cm}(E)}, 
\end{equation}
\begin{equation}
\mathcal{K}^B_{dncm}(E) = -(v_f/|E|)^2\mathcal{K}^A_{dncm}(E),
\end{equation}
\begin{equation} 
\tilde{\rho}^{Ay}_{dncm}(y;E) =  \sigma_p(E)F^c_{dncm}(y-y_i;E) - (1/\gamma +\sigma_e(E))F^s_{dncm}(y-y_i;E),
\end{equation}
and
\begin{equation}
\begin{split}
\tilde{\rho}^{By}_{dncm}(y;E)
&= \left\{ \sigma_p(E)(-k_nk_m + \kappa_n(E)\kappa_m(E)) \right. \\
&  \quad\left. +(1/\gamma+\sigma_e(E))\sgn{E}\sgn{y-y_i}(k_n\kappa_m(E)+k_m\kappa_n(E))\right\}F^c_{dncm}(y-y_i;E) \\
&+ \left\{ \sigma_p(E)\sgn{E}\sgn{y-y_i}(k_n \kappa_m(E)+k_m\kappa_n(E))\right. \\
& \quad\left. - (1/\gamma+\sigma_e(E))\sgn{E}\sgn{y-y_i}(k_n\kappa_m(E)+k_m\kappa_n(E))\right\}F^s_{dncm}(y-y_i;E)
\end{split}
\end{equation}
where
\begin{equation}
F^c_{dncm}(y;E) = \cos\left[\text{sgn}(E)(\kappa_{dn}(E)+\kappa_{cm}(E))|y| + (K_{dy}-K_{cy})y\right]
\end{equation}
and
\begin{equation}
F^s_{dncm}(y;E) = \sin\left[\text{sgn}(E)(\kappa_{dn}(E)+\kappa_{cm}(E))|y| + (K_{dy}-K_{cy})y\right].
\end{equation}
The Fourier transform of each component is carried out exactly as for the 2DEG, using Eq. (\ref{eq:fourier}), and the results for the AGNR
are shown in Section~\ref{Ch_armchair}.

\bibliographystyle{apsrev4-1}

%


\begin{thebibliography}{32}%
\makeatletter
\providecommand \@ifxundefined [1]{%
 \@ifx{#1\undefined}
}%
\providecommand \@ifnum [1]{%
 \ifnum #1\expandafter \@firstoftwo
 \else \expandafter \@secondoftwo
 \fi
}%
\providecommand \@ifx [1]{%
 \ifx #1\expandafter \@firstoftwo
 \else \expandafter \@secondoftwo
 \fi
}%
\providecommand \natexlab [1]{#1}%
\providecommand \enquote  [1]{``#1''}%
\providecommand \bibnamefont  [1]{#1}%
\providecommand \bibfnamefont [1]{#1}%
\providecommand \citenamefont [1]{#1}%
\providecommand \href@noop [0]{\@secondoftwo}%
\providecommand \href [0]{\begingroup \@sanitize@url \@href}%
\providecommand \@href[1]{\@@startlink{#1}\@@href}%
\providecommand \@@href[1]{\endgroup#1\@@endlink}%
\providecommand \@sanitize@url [0]{\catcode `\\12\catcode `\$12\catcode
  `\&12\catcode `\#12\catcode `\^12\catcode `\_12\catcode `\%12\relax}%
\providecommand \@@startlink[1]{}%
\providecommand \@@endlink[0]{}%
\providecommand \url  [0]{\begingroup\@sanitize@url \@url }%
\providecommand \@url [1]{\endgroup\@href {#1}{\urlprefix }}%
\providecommand \urlprefix  [0]{URL }%
\providecommand \Eprint [0]{\href }%
\providecommand \doibase [0]{http://dx.doi.org/}%
\providecommand \selectlanguage [0]{\@gobble}%
\providecommand \bibinfo  [0]{\@secondoftwo}%
\providecommand \bibfield  [0]{\@secondoftwo}%
\providecommand \translation [1]{[#1]}%
\providecommand \BibitemOpen [0]{}%
\providecommand \bibitemStop [0]{}%
\providecommand \bibitemNoStop [0]{.\EOS\space}%
\providecommand \EOS [0]{\spacefactor3000\relax}%
\providecommand \BibitemShut  [1]{\csname bibitem#1\endcsname}%
\let\auto@bib@innerbib\@empty
\bibitem [{\citenamefont {Castro~Neto}\ \emph {et~al.}(2009)\citenamefont
  {Castro~Neto}, \citenamefont {Guinea}, \citenamefont {Peres}, \citenamefont
  {Novoselov},\ and\ \citenamefont {Geim}}]{CastroNeto:2009cl}%
  \BibitemOpen
  \bibfield  {author} {\bibinfo {author} {\bibfnamefont {A.~H.}\ \bibnamefont
  {Castro~Neto}}, \bibinfo {author} {\bibfnamefont {F.}~\bibnamefont {Guinea}},
  \bibinfo {author} {\bibfnamefont {N.~M.~R.}\ \bibnamefont {Peres}}, \bibinfo
  {author} {\bibfnamefont {K.~S.}\ \bibnamefont {Novoselov}}, \ and\ \bibinfo
  {author} {\bibfnamefont {A.~K.}\ \bibnamefont {Geim}},\ }\href@noop {}
  {\bibfield  {journal} {\bibinfo  {journal} {Reviews Of Modern Physics}\
  }\textbf {\bibinfo {volume} {81}},\ \bibinfo {pages} {109} (\bibinfo {year}
  {2009})}\BibitemShut {NoStop}%
\bibitem [{\citenamefont {Schedin}\ \emph {et~al.}(2007)\citenamefont
  {Schedin}, \citenamefont {Geim}, \citenamefont {Morozov}, \citenamefont
  {Hill}, \citenamefont {Blake}, \citenamefont {Katsnelson},\ and\
  \citenamefont {Novoselov}}]{Schedin:2007hs}%
  \BibitemOpen
  \bibfield  {author} {\bibinfo {author} {\bibfnamefont {F.}~\bibnamefont
  {Schedin}}, \bibinfo {author} {\bibfnamefont {A.~K.}\ \bibnamefont {Geim}},
  \bibinfo {author} {\bibfnamefont {S.~V.}\ \bibnamefont {Morozov}}, \bibinfo
  {author} {\bibfnamefont {E.~W.}\ \bibnamefont {Hill}}, \bibinfo {author}
  {\bibfnamefont {P.}~\bibnamefont {Blake}}, \bibinfo {author} {\bibfnamefont
  {M.~I.}\ \bibnamefont {Katsnelson}}, \ and\ \bibinfo {author} {\bibfnamefont
  {K.~S.}\ \bibnamefont {Novoselov}},\ }\href@noop {} {\bibfield  {journal}
  {\bibinfo  {journal} {Nature Materials}\ }\textbf {\bibinfo {volume} {6}},\
  \bibinfo {pages} {652} (\bibinfo {year} {2007})}\BibitemShut {NoStop}%
\bibitem [{\citenamefont {Brar}\ \emph {et~al.}(2011)\citenamefont {Brar},
  \citenamefont {Decker}, \citenamefont {Solowan}, \citenamefont {Wang},
  \citenamefont {Maserati}, \citenamefont {Chan}, \citenamefont {Lee},
  \citenamefont {Girit}, \citenamefont {Zettl}, \citenamefont {Louie},
  \citenamefont {Cohen},\ and\ \citenamefont {Crommie}}]{Brar:2011iq}%
  \BibitemOpen
  \bibfield  {author} {\bibinfo {author} {\bibfnamefont {V.~W.}\ \bibnamefont
  {Brar}}, \bibinfo {author} {\bibfnamefont {R.}~\bibnamefont {Decker}},
  \bibinfo {author} {\bibfnamefont {H.-M.}\ \bibnamefont {Solowan}}, \bibinfo
  {author} {\bibfnamefont {Y.}~\bibnamefont {Wang}}, \bibinfo {author}
  {\bibfnamefont {L.}~\bibnamefont {Maserati}}, \bibinfo {author}
  {\bibfnamefont {K.~T.}\ \bibnamefont {Chan}}, \bibinfo {author}
  {\bibfnamefont {H.}~\bibnamefont {Lee}}, \bibinfo {author} {\bibfnamefont
  {{\c C}.~O.}\ \bibnamefont {Girit}}, \bibinfo {author} {\bibfnamefont
  {A.}~\bibnamefont {Zettl}}, \bibinfo {author} {\bibfnamefont {S.~G.}\
  \bibnamefont {Louie}}, \bibinfo {author} {\bibfnamefont {M.~L.}\ \bibnamefont
  {Cohen}}, \ and\ \bibinfo {author} {\bibfnamefont {M.~F.}\ \bibnamefont
  {Crommie}},\ }\href@noop {} {\bibfield  {journal} {\bibinfo  {journal}
  {Nature Physics}\ }\textbf {\bibinfo {volume} {7}},\ \bibinfo {pages} {43}
  (\bibinfo {year} {2011})}\BibitemShut {NoStop}%
\bibitem [{\citenamefont {Peres}(2010)}]{Peres:2010tn}%
  \BibitemOpen
  \bibfield  {author} {\bibinfo {author} {\bibfnamefont {N.}~\bibnamefont
  {Peres}},\ }\href@noop {} {\bibfield  {journal} {\bibinfo  {journal} {Reviews
  Of Modern Physics}\ }\textbf {\bibinfo {volume} {82}},\ \bibinfo {pages}
  {2673} (\bibinfo {year} {2010})}\BibitemShut {NoStop}%
\bibitem [{\citenamefont {Connolly}\ and\ \citenamefont
  {Smith}(2010)}]{Connolly:2010bc}%
  \BibitemOpen
  \bibfield  {author} {\bibinfo {author} {\bibfnamefont {M.~R.}\ \bibnamefont
  {Connolly}}\ and\ \bibinfo {author} {\bibfnamefont {C.~G.}\ \bibnamefont
  {Smith}},\ }\href@noop {} {\bibfield  {journal} {\bibinfo  {journal}
  {Philosophical Transactions Of The Royal Society A-Mathematical Physical And
  Engineering Sciences}\ }\textbf {\bibinfo {volume} {368}},\ \bibinfo {pages}
  {5379} (\bibinfo {year} {2010})}\BibitemShut {NoStop}%
\bibitem [{\citenamefont {Deshpande}\ and\ \citenamefont
  {LeRoy}(2012)}]{Deshpande:2012jg}%
  \BibitemOpen
  \bibfield  {author} {\bibinfo {author} {\bibfnamefont {A.}~\bibnamefont
  {Deshpande}}\ and\ \bibinfo {author} {\bibfnamefont {B.~J.}\ \bibnamefont
  {LeRoy}},\ }\href@noop {} {\bibfield  {journal} {\bibinfo  {journal} {Physica
  E-Low-Dimensional Systems {\&} Nanostructures}\ }\textbf {\bibinfo {volume}
  {44}},\ \bibinfo {pages} {743} (\bibinfo {year} {2012})}\BibitemShut
  {NoStop}%
\bibitem [{\citenamefont {Rutter}\ \emph {et~al.}(2007)\citenamefont {Rutter},
  \citenamefont {Crain}, \citenamefont {Guisinger}, \citenamefont {Li},
  \citenamefont {First},\ and\ \citenamefont {Stroscio}}]{Rutter:2007ep}%
  \BibitemOpen
  \bibfield  {author} {\bibinfo {author} {\bibfnamefont {G.~M.}\ \bibnamefont
  {Rutter}}, \bibinfo {author} {\bibfnamefont {J.~N.}\ \bibnamefont {Crain}},
  \bibinfo {author} {\bibfnamefont {N.~P.}\ \bibnamefont {Guisinger}}, \bibinfo
  {author} {\bibfnamefont {T.}~\bibnamefont {Li}}, \bibinfo {author}
  {\bibfnamefont {P.~N.}\ \bibnamefont {First}}, \ and\ \bibinfo {author}
  {\bibfnamefont {J.~A.}\ \bibnamefont {Stroscio}},\ }\href@noop {} {\bibfield
  {journal} {\bibinfo  {journal} {Science (New York, NY)}\ }\textbf {\bibinfo
  {volume} {317}},\ \bibinfo {pages} {219} (\bibinfo {year}
  {2007})}\BibitemShut {NoStop}%
\bibitem [{\citenamefont {Mallet}\ \emph {et~al.}(2007)\citenamefont {Mallet},
  \citenamefont {Varchon}, \citenamefont {Naud}, \citenamefont {Magaud},
  \citenamefont {Berger},\ and\ \citenamefont {Veuillen}}]{Mallet:2007fg}%
  \BibitemOpen
  \bibfield  {author} {\bibinfo {author} {\bibfnamefont {P.}~\bibnamefont
  {Mallet}}, \bibinfo {author} {\bibfnamefont {F.}~\bibnamefont {Varchon}},
  \bibinfo {author} {\bibfnamefont {C.}~\bibnamefont {Naud}}, \bibinfo {author}
  {\bibfnamefont {L.}~\bibnamefont {Magaud}}, \bibinfo {author} {\bibfnamefont
  {C.}~\bibnamefont {Berger}}, \ and\ \bibinfo {author} {\bibfnamefont {J.-Y.}\
  \bibnamefont {Veuillen}},\ }\href@noop {} {\bibfield  {journal} {\bibinfo
  {journal} {Physical Review B}\ }\textbf {\bibinfo {volume} {76}},\  (\bibinfo
  {year} {2007})}\BibitemShut {NoStop}%
\bibitem [{\citenamefont {Xue}\ \emph {et~al.}(2012)\citenamefont {Xue},
  \citenamefont {Sanchez-Yamagishi}, \citenamefont {Watanabe}, \citenamefont
  {Taniguchi}, \citenamefont {Jarillo-Herrero},\ and\ \citenamefont
  {LeRoy}}]{Xue:2012hd}%
  \BibitemOpen
  \bibfield  {author} {\bibinfo {author} {\bibfnamefont {J.}~\bibnamefont
  {Xue}}, \bibinfo {author} {\bibfnamefont {J.}~\bibnamefont
  {Sanchez-Yamagishi}}, \bibinfo {author} {\bibfnamefont {K.}~\bibnamefont
  {Watanabe}}, \bibinfo {author} {\bibfnamefont {T.}~\bibnamefont {Taniguchi}},
  \bibinfo {author} {\bibfnamefont {P.}~\bibnamefont {Jarillo-Herrero}}, \ and\
  \bibinfo {author} {\bibfnamefont {B.}~\bibnamefont {LeRoy}},\ }\href@noop {}
  {\bibfield  {journal} {\bibinfo  {journal} {Physical Review Letters}\
  }\textbf {\bibinfo {volume} {108}} (\bibinfo {year} {2012})}\BibitemShut
  {NoStop}%
\bibitem [{\citenamefont {Zhang}\ \emph {et~al.}(2009)\citenamefont {Zhang},
  \citenamefont {Brar}, \citenamefont {Girit}, \citenamefont {Zettl},\ and\
  \citenamefont {Crommie}}]{Zhang:2009ce}%
  \BibitemOpen
  \bibfield  {author} {\bibinfo {author} {\bibfnamefont {Y.}~\bibnamefont
  {Zhang}}, \bibinfo {author} {\bibfnamefont {V.~W.}\ \bibnamefont {Brar}},
  \bibinfo {author} {\bibfnamefont {C.}~\bibnamefont {Girit}}, \bibinfo
  {author} {\bibfnamefont {A.}~\bibnamefont {Zettl}}, \ and\ \bibinfo {author}
  {\bibfnamefont {M.~F.}\ \bibnamefont {Crommie}},\ }\href@noop {} {\bibfield
  {journal} {\bibinfo  {journal} {Nature Physics}\ }\textbf {\bibinfo {volume}
  {5}},\ \bibinfo {pages} {722} (\bibinfo {year} {2009})}\BibitemShut {NoStop}%
\bibitem [{\citenamefont {Ji}\ \emph {et~al.}(2011)\citenamefont {Ji},
  \citenamefont {Hannon}, \citenamefont {Tromp}, \citenamefont {Perebeinos},
  \citenamefont {Tersoff},\ and\ \citenamefont {Ross}}]{Ji:2011bla}%
  \BibitemOpen
  \bibfield  {author} {\bibinfo {author} {\bibfnamefont {S.-H.}\ \bibnamefont
  {Ji}}, \bibinfo {author} {\bibfnamefont {J.~B.}\ \bibnamefont {Hannon}},
  \bibinfo {author} {\bibfnamefont {R.~M.}\ \bibnamefont {Tromp}}, \bibinfo
  {author} {\bibfnamefont {V.}~\bibnamefont {Perebeinos}}, \bibinfo {author}
  {\bibfnamefont {J.}~\bibnamefont {Tersoff}}, \ and\ \bibinfo {author}
  {\bibfnamefont {F.~M.}\ \bibnamefont {Ross}},\ }\href@noop {} {\bibfield
  {journal} {\bibinfo  {journal} {Nature Materials}\ }\textbf {\bibinfo
  {volume} {11}},\ \bibinfo {pages} {114} (\bibinfo {year} {2011})}\BibitemShut
  {NoStop}%
\bibitem [{\citenamefont {Wang}\ \emph {et~al.}(2012)\citenamefont {Wang},
  \citenamefont {Munakata},\ and\ \citenamefont {Rozler}}]{Wang:2012tj}%
  \BibitemOpen
  \bibfield  {author} {\bibinfo {author} {\bibfnamefont {W.}~\bibnamefont
  {Wang}}, \bibinfo {author} {\bibfnamefont {K.}~\bibnamefont {Munakata}}, \
  and\ \bibinfo {author} {\bibfnamefont {M.}~\bibnamefont {Rozler}},\
  }\href@noop {} {\bibfield  {journal} {\bibinfo  {journal} {arXiv.org}\ }
  (\bibinfo {year} {2012})}\BibitemShut {NoStop}%
\bibitem [{\citenamefont {Giannazzo}\ \emph {et~al.}(2012)\citenamefont
  {Giannazzo}, \citenamefont {Deretzis}, \citenamefont {La~Magna},
  \citenamefont {Roccaforte},\ and\ \citenamefont
  {Yakimova}}]{Giannazzo:2012ib}%
  \BibitemOpen
  \bibfield  {author} {\bibinfo {author} {\bibfnamefont {F.}~\bibnamefont
  {Giannazzo}}, \bibinfo {author} {\bibfnamefont {I.}~\bibnamefont {Deretzis}},
  \bibinfo {author} {\bibfnamefont {A.}~\bibnamefont {La~Magna}}, \bibinfo
  {author} {\bibfnamefont {F.}~\bibnamefont {Roccaforte}}, \ and\ \bibinfo
  {author} {\bibfnamefont {R.}~\bibnamefont {Yakimova}},\ }\href@noop {}
  {\bibfield  {journal} {\bibinfo  {journal} {Physical Review B}\ }\textbf
  {\bibinfo {volume} {86}},\ \bibinfo {pages} {235422} (\bibinfo {year}
  {2012})}\BibitemShut {NoStop}%
\bibitem [{\citenamefont {Han}\ \emph {et~al.}(2007)\citenamefont {Han},
  \citenamefont {Ozyilmaz}, \citenamefont {Zhang},\ and\ \citenamefont
  {Kim}}]{Han:2007bl}%
  \BibitemOpen
  \bibfield  {author} {\bibinfo {author} {\bibfnamefont {M.}~\bibnamefont
  {Han}}, \bibinfo {author} {\bibfnamefont {B.}~\bibnamefont {Ozyilmaz}},
  \bibinfo {author} {\bibfnamefont {Y.}~\bibnamefont {Zhang}}, \ and\ \bibinfo
  {author} {\bibfnamefont {P.}~\bibnamefont {Kim}},\ }\href@noop {} {\bibfield
  {journal} {\bibinfo  {journal} {Physical Review Letters}\ }\textbf {\bibinfo
  {volume} {98}},\ \bibinfo {pages} {206805} (\bibinfo {year}
  {2007})}\BibitemShut {NoStop}%
\bibitem [{\citenamefont {Biro}\ and\ \citenamefont
  {Lambin}(2010)}]{Biro:2010ez}%
  \BibitemOpen
  \bibfield  {author} {\bibinfo {author} {\bibfnamefont {L.~P.}\ \bibnamefont
  {Biro}}\ and\ \bibinfo {author} {\bibfnamefont {P.}~\bibnamefont {Lambin}},\
  }\href@noop {} {\bibfield  {journal} {\bibinfo  {journal} {Carbon}\ }\textbf
  {\bibinfo {volume} {48}},\ \bibinfo {pages} {2677} (\bibinfo {year}
  {2010})}\BibitemShut {NoStop}%
\bibitem [{\citenamefont {Campos}\ \emph {et~al.}(2009)\citenamefont {Campos},
  \citenamefont {Manfrinato}, \citenamefont {Sanchez-Yamagishi}, \citenamefont
  {Kong},\ and\ \citenamefont {Jarillo-Herrero}}]{Campos:2009cd}%
  \BibitemOpen
  \bibfield  {author} {\bibinfo {author} {\bibfnamefont {L.~C.}\ \bibnamefont
  {Campos}}, \bibinfo {author} {\bibfnamefont {V.~R.}\ \bibnamefont
  {Manfrinato}}, \bibinfo {author} {\bibfnamefont {J.~D.}\ \bibnamefont
  {Sanchez-Yamagishi}}, \bibinfo {author} {\bibfnamefont {J.}~\bibnamefont
  {Kong}}, \ and\ \bibinfo {author} {\bibfnamefont {P.}~\bibnamefont
  {Jarillo-Herrero}},\ }\href@noop {} {\bibfield  {journal} {\bibinfo
  {journal} {Nano Letters}\ }\textbf {\bibinfo {volume} {9}},\ \bibinfo {pages}
  {2600} (\bibinfo {year} {2009})}\BibitemShut {NoStop}%
\bibitem [{\citenamefont {Girit}\ \emph {et~al.}(2009)\citenamefont {Girit},
  \citenamefont {Meyer}, \citenamefont {Erni}, \citenamefont {Rossell},
  \citenamefont {Kisielowski}, \citenamefont {Yang}, \citenamefont {Park},
  \citenamefont {Crommie}, \citenamefont {Cohen},\ and\ \citenamefont
  {Louie}}]{Girit:2009tz}%
  \BibitemOpen
  \bibfield  {author} {\bibinfo {author} {\bibfnamefont {{\c C}.~{\"O}.}\
  \bibnamefont {Girit}}, \bibinfo {author} {\bibfnamefont {J.~C.}\ \bibnamefont
  {Meyer}}, \bibinfo {author} {\bibfnamefont {R.}~\bibnamefont {Erni}},
  \bibinfo {author} {\bibfnamefont {M.~D.}\ \bibnamefont {Rossell}}, \bibinfo
  {author} {\bibfnamefont {C.}~\bibnamefont {Kisielowski}}, \bibinfo {author}
  {\bibfnamefont {L.}~\bibnamefont {Yang}}, \bibinfo {author} {\bibfnamefont
  {C.~H.}\ \bibnamefont {Park}}, \bibinfo {author} {\bibfnamefont {M.~F.}\
  \bibnamefont {Crommie}}, \bibinfo {author} {\bibfnamefont {M.~L.}\
  \bibnamefont {Cohen}}, \ and\ \bibinfo {author} {\bibfnamefont {S.~G.}\
  \bibnamefont {Louie}},\ }\href@noop {} {\bibfield  {journal} {\bibinfo
  {journal} {Science (New York, NY)}\ }\textbf {\bibinfo {volume} {323}},\
  \bibinfo {pages} {1705} (\bibinfo {year} {2009})}\BibitemShut {NoStop}%
\bibitem [{\citenamefont {Cai}\ \emph {et~al.}(2010)\citenamefont {Cai},
  \citenamefont {Ruffieux}, \citenamefont {Jaafar}, \citenamefont {Bieri},
  \citenamefont {Braun}, \citenamefont {Blankenburg}, \citenamefont {Muoth},
  \citenamefont {Seitsonen}, \citenamefont {Saleh}, \citenamefont {Feng},
  \citenamefont {M{\"u}llen},\ and\ \citenamefont {Fasel}}]{Cai:2010bd}%
  \BibitemOpen
  \bibfield  {author} {\bibinfo {author} {\bibfnamefont {J.}~\bibnamefont
  {Cai}}, \bibinfo {author} {\bibfnamefont {P.}~\bibnamefont {Ruffieux}},
  \bibinfo {author} {\bibfnamefont {R.}~\bibnamefont {Jaafar}}, \bibinfo
  {author} {\bibfnamefont {M.}~\bibnamefont {Bieri}}, \bibinfo {author}
  {\bibfnamefont {T.}~\bibnamefont {Braun}}, \bibinfo {author} {\bibfnamefont
  {S.}~\bibnamefont {Blankenburg}}, \bibinfo {author} {\bibfnamefont
  {M.}~\bibnamefont {Muoth}}, \bibinfo {author} {\bibfnamefont {A.~P.}\
  \bibnamefont {Seitsonen}}, \bibinfo {author} {\bibfnamefont {M.}~\bibnamefont
  {Saleh}}, \bibinfo {author} {\bibfnamefont {X.}~\bibnamefont {Feng}},
  \bibinfo {author} {\bibfnamefont {K.}~\bibnamefont {M{\"u}llen}}, \ and\
  \bibinfo {author} {\bibfnamefont {R.}~\bibnamefont {Fasel}},\ }\href@noop {}
  {\bibfield  {journal} {\bibinfo  {journal} {Nature}\ }\textbf {\bibinfo
  {volume} {466}},\ \bibinfo {pages} {470} (\bibinfo {year}
  {2010})}\BibitemShut {NoStop}%
\bibitem [{\citenamefont {Li}\ \emph {et~al.}(2008)\citenamefont {Li},
  \citenamefont {Wang}, \citenamefont {Zhang}, \citenamefont {Lee},\ and\
  \citenamefont {Dai}}]{Li:2008ht}%
  \BibitemOpen
  \bibfield  {author} {\bibinfo {author} {\bibfnamefont {X.}~\bibnamefont
  {Li}}, \bibinfo {author} {\bibfnamefont {X.}~\bibnamefont {Wang}}, \bibinfo
  {author} {\bibfnamefont {L.}~\bibnamefont {Zhang}}, \bibinfo {author}
  {\bibfnamefont {S.}~\bibnamefont {Lee}}, \ and\ \bibinfo {author}
  {\bibfnamefont {H.}~\bibnamefont {Dai}},\ }\href@noop {} {\bibfield
  {journal} {\bibinfo  {journal} {Science (New York, NY)}\ }\textbf {\bibinfo
  {volume} {319}},\ \bibinfo {pages} {1229} (\bibinfo {year}
  {2008})}\BibitemShut {NoStop}%
\bibitem [{\citenamefont {Fujita}\ \emph {et~al.}(1996)\citenamefont {Fujita},
  \citenamefont {Wakabayashi}, \citenamefont {Nakada},\ and\ \citenamefont
  {Kusakabe}}]{Fujita:1996vs}%
  \BibitemOpen
  \bibfield  {author} {\bibinfo {author} {\bibfnamefont {M.}~\bibnamefont
  {Fujita}}, \bibinfo {author} {\bibfnamefont {K.}~\bibnamefont {Wakabayashi}},
  \bibinfo {author} {\bibfnamefont {K.}~\bibnamefont {Nakada}}, \ and\ \bibinfo
  {author} {\bibfnamefont {K.}~\bibnamefont {Kusakabe}},\ }\href@noop {}
  {\bibfield  {journal} {\bibinfo  {journal} {Journal Of The Physical Society
  Of Japan}\ }\textbf {\bibinfo {volume} {65}},\ \bibinfo {pages} {1920}
  (\bibinfo {year} {1996})}\BibitemShut {NoStop}%
\bibitem [{\citenamefont {Nakada}\ \emph {et~al.}(1996)\citenamefont {Nakada},
  \citenamefont {Fujita}, \citenamefont {Dresselhaus},\ and\ \citenamefont
  {Dresselhaus}}]{Nakada:1996us}%
  \BibitemOpen
  \bibfield  {author} {\bibinfo {author} {\bibfnamefont {K.}~\bibnamefont
  {Nakada}}, \bibinfo {author} {\bibfnamefont {M.}~\bibnamefont {Fujita}},
  \bibinfo {author} {\bibfnamefont {G.}~\bibnamefont {Dresselhaus}}, \ and\
  \bibinfo {author} {\bibfnamefont {M.}~\bibnamefont {Dresselhaus}},\
  }\href@noop {} {\bibfield  {journal} {\bibinfo  {journal} {Physical Review
  B}\ }\textbf {\bibinfo {volume} {54}},\ \bibinfo {pages} {17954} (\bibinfo
  {year} {1996})}\BibitemShut {NoStop}%
\bibitem [{\citenamefont {Tao}\ \emph {et~al.}(2011)\citenamefont {Tao},
  \citenamefont {Jiao}, \citenamefont {Yazyev}, \citenamefont {Chen},
  \citenamefont {Feng}, \citenamefont {Zhang}, \citenamefont {Capaz},
  \citenamefont {Tour}, \citenamefont {Zettl}, \citenamefont {Louie},
  \citenamefont {Dai},\ and\ \citenamefont {Crommie}}]{Tao:2011bk}%
  \BibitemOpen
  \bibfield  {author} {\bibinfo {author} {\bibfnamefont {C.}~\bibnamefont
  {Tao}}, \bibinfo {author} {\bibfnamefont {L.}~\bibnamefont {Jiao}}, \bibinfo
  {author} {\bibfnamefont {O.~V.}\ \bibnamefont {Yazyev}}, \bibinfo {author}
  {\bibfnamefont {Y.-C.}\ \bibnamefont {Chen}}, \bibinfo {author}
  {\bibfnamefont {J.}~\bibnamefont {Feng}}, \bibinfo {author} {\bibfnamefont
  {X.}~\bibnamefont {Zhang}}, \bibinfo {author} {\bibfnamefont {R.~B.}\
  \bibnamefont {Capaz}}, \bibinfo {author} {\bibfnamefont {J.~M.}\ \bibnamefont
  {Tour}}, \bibinfo {author} {\bibfnamefont {A.}~\bibnamefont {Zettl}},
  \bibinfo {author} {\bibfnamefont {S.~G.}\ \bibnamefont {Louie}}, \bibinfo
  {author} {\bibfnamefont {H.}~\bibnamefont {Dai}}, \ and\ \bibinfo {author}
  {\bibfnamefont {M.~F.}\ \bibnamefont {Crommie}},\ }\href@noop {} {\bibfield
  {journal} {\bibinfo  {journal} {Nature Physics}\ }\textbf {\bibinfo {volume}
  {7}},\ \bibinfo {pages} {616} (\bibinfo {year} {2011})}\BibitemShut {NoStop}%
\bibitem [{\citenamefont {Palacios}\ \emph {et~al.}(2010)\citenamefont
  {Palacios}, \citenamefont {Fernandez-Rossier}, \citenamefont {Brey},\ and\
  \citenamefont {Fertig}}]{Palacios:2010ep}%
  \BibitemOpen
  \bibfield  {author} {\bibinfo {author} {\bibfnamefont {J.~J.}\ \bibnamefont
  {Palacios}}, \bibinfo {author} {\bibfnamefont {J.}~\bibnamefont
  {Fernandez-Rossier}}, \bibinfo {author} {\bibfnamefont {L.}~\bibnamefont
  {Brey}}, \ and\ \bibinfo {author} {\bibfnamefont {H.~A.}\ \bibnamefont
  {Fertig}},\ }\href@noop {} {\bibfield  {journal} {\bibinfo  {journal}
  {Semiconductor Science And Technology}\ }\textbf {\bibinfo {volume} {25}},\
  \bibinfo {pages} {033003} (\bibinfo {year} {2010})}\BibitemShut {NoStop}%
\bibitem{Raza}
{\it Graphene Nanoelectronics; Metrology, Synthesis,  Properties and Applications},
H. Raza, Editor, Springer-Verlag, Berlin Heidelberg 2012.
\bibitem [{\citenamefont {Pereg-Barnea}\ and\ \citenamefont
  {Macdonald}(2008)}]{PeregBarnea:2008ig}%
  \BibitemOpen
  \bibfield  {author} {\bibinfo {author} {\bibfnamefont {T.}~\bibnamefont
  {Pereg-Barnea}}\ and\ \bibinfo {author} {\bibfnamefont {A.}~\bibnamefont
  {Macdonald}},\ }\href@noop {} {\bibfield  {journal} {\bibinfo  {journal}
  {Physical Review B}\ }\textbf {\bibinfo {volume} {78}},\ \bibinfo {pages}
  {014201} (\bibinfo {year} {2008})}\BibitemShut {NoStop}%
\bibitem [{\citenamefont {Bena}(2008)}]{Bena:2008iw}%
  \BibitemOpen
  \bibfield  {author} {\bibinfo {author} {\bibfnamefont {C.}~\bibnamefont
  {Bena}},\ }\href@noop {} {\bibfield  {journal} {\bibinfo  {journal} {Physical
  Review Letters}\ }\textbf {\bibinfo {volume} {100}},\ \bibinfo {pages}
  {076601} (\bibinfo {year} {2008})}\BibitemShut {NoStop}%
\bibitem [{\citenamefont {Petersen}\ \emph {et~al.}(2000)\citenamefont
  {Petersen}, \citenamefont {Hofmann}, \citenamefont {Plummer},\ and\
  \citenamefont {Besenbacher}}]{Anonymous:IEkMmlpc}%
  \BibitemOpen
  \bibfield  {author} {\bibinfo {author} {\bibfnamefont {L.}~\bibnamefont
  {Petersen}}, \bibinfo {author} {\bibfnamefont {P.}~\bibnamefont {Hofmann}},
  \bibinfo {author} {\bibfnamefont {E.~W.}\ \bibnamefont {Plummer}}, \ and\
  \bibinfo {author} {\bibfnamefont {F.}~\bibnamefont {Besenbacher}},\
  }\href@noop {} {\bibfield  {journal} {\bibinfo  {journal} {Journal of
  electron spectroscopy and related phenomena}\ }\textbf {\bibinfo {volume}
  {109}},\ \bibinfo {pages} {97} (\bibinfo {year} {2000})}\BibitemShut
  {NoStop}%
\bibitem [{\citenamefont {Balatsky}\ \emph {et~al.}(2006)\citenamefont
  {Balatsky}, \citenamefont {Vekhter},\ and\ \citenamefont
  {Zhu}}]{Balatsky:2006ce}%
  \BibitemOpen
  \bibfield  {author} {\bibinfo {author} {\bibfnamefont {A.~V.}\ \bibnamefont
  {Balatsky}}, \bibinfo {author} {\bibfnamefont {I.}~\bibnamefont {Vekhter}}, \
  and\ \bibinfo {author} {\bibfnamefont {J.-X.}\ \bibnamefont {Zhu}},\
  }\href@noop {} {\bibfield  {journal} {\bibinfo  {journal} {Reviews Of Modern
  Physics}\ }\textbf {\bibinfo {volume} {78}},\ \bibinfo {pages} {373}
  (\bibinfo {year} {2006})}\BibitemShut {NoStop}%
\bibitem [{\citenamefont {Boese}\ \emph {et~al.}(2000)\citenamefont {Boese},
  \citenamefont {Lischka},\ and\ \citenamefont {Reichl}}]{2000PhRvB..61.5632B}%
  \BibitemOpen
  \bibfield  {author} {\bibinfo {author} {\bibfnamefont {D.}~\bibnamefont
  {Boese}}, \bibinfo {author} {\bibfnamefont {M.}~\bibnamefont {Lischka}}, \
  and\ \bibinfo {author} {\bibfnamefont {L.~E.}\ \bibnamefont {Reichl}},\
  }\href@noop {} {\bibfield  {journal} {\bibinfo  {journal} {Physical Review B
  (Condensed Matter and Materials Physics)}\ }\textbf {\bibinfo {volume}
  {61}},\ \bibinfo {pages} {5632} (\bibinfo {year} {2000})}\BibitemShut
  {NoStop}%
\bibitem [{\citenamefont {Kazymyrenko}\ and\ \citenamefont
  {Waintal}(2008)}]{Kazymyrenko:2008hk}%
  \BibitemOpen
  \bibfield  {author} {\bibinfo {author} {\bibfnamefont {K.}~\bibnamefont
  {Kazymyrenko}}\ and\ \bibinfo {author} {\bibfnamefont {X.}~\bibnamefont
  {Waintal}},\ }\href@noop {} {\bibfield  {journal} {\bibinfo  {journal}
  {Physical Review B}\ }\textbf {\bibinfo {volume} {77}},\ \bibinfo {pages}
  {115119} (\bibinfo {year} {2008})}\BibitemShut {NoStop}%
\bibitem [{\citenamefont {Wakabayashi}\ \emph {et~al.}(2009)\citenamefont
  {Wakabayashi}, \citenamefont {Takane}, \citenamefont {Yamamoto},\ and\
  \citenamefont {Sigrist}}]{Wakabayashi:2009dh}%
  \BibitemOpen
  \bibfield  {author} {\bibinfo {author} {\bibfnamefont {K.}~\bibnamefont
  {Wakabayashi}}, \bibinfo {author} {\bibfnamefont {Y.}~\bibnamefont {Takane}},
  \bibinfo {author} {\bibfnamefont {M.}~\bibnamefont {Yamamoto}}, \ and\
  \bibinfo {author} {\bibfnamefont {M.}~\bibnamefont {Sigrist}},\ }\href@noop
  {} {\bibfield  {journal} {\bibinfo  {journal} {New Journal Of Physics}\
  }\textbf {\bibinfo {volume} {11}},\ \bibinfo {pages} {095016} (\bibinfo
  {year} {2009})}\BibitemShut {NoStop}%
\bibitem{Eco}
E. N. Economou, {\it Green's Functions in Quantum Physics, 3rd Ed.}, Springer Verlag, Berlin 2006
\bibitem [{\citenamefont {Bagwell}(1990)}]{Bagwell:2012te}%
  \BibitemOpen
  \bibfield  {author} {\bibinfo {author} {\bibfnamefont {P.}~\bibnamefont
  {Bagwell}},\ }\href@noop {} {\bibfield  {journal} {\bibinfo  {journal}
  {Journal Of Physics-Condensed Matter}\ }\textbf {\bibinfo {volume} {2}},\
  \bibinfo {pages} {6179} (\bibinfo {year} {1990})}\BibitemShut {NoStop}%
\bibitem [{\citenamefont {Brey}\ and\ \citenamefont
  {Fertig}(2006)}]{Brey:2006cb}%
  \BibitemOpen
  \bibfield  {author} {\bibinfo {author} {\bibfnamefont {L.}~\bibnamefont
  {Brey}}\ and\ \bibinfo {author} {\bibfnamefont {H.}~\bibnamefont {Fertig}},\
  }\href@noop {} {\bibfield  {journal} {\bibinfo  {journal} {Physical Review
  B}\ }\textbf {\bibinfo {volume} {73}},\ \bibinfo {pages} {235411} (\bibinfo
  {year} {2006})}\BibitemShut {NoStop}%
\end{thebibliography}

\end{document}